\documentclass{JHEP3}

\usepackage{graphicx}
\usepackage{amssymb}
\usepackage[mathscr]{eucal}
\usepackage{amsmath}
\usepackage{cite}
\usepackage{afterpage}

\newcommand{\gsimm}{\raise.3ex\hbox{$>$\kern-.75em\lower1ex\hbox{$\sim$}}}
\newcommand{\lsimm}{\raise.3ex\hbox{$<$\kern-.75em\lower1ex\hbox{$\sim$}}}

\title{Gif Lectures on Cosmic Acceleration\footnote{ Ecole de Gif, Batz sur mer, 21-25 September 2009.}}

\author{Philippe Brax \\
  Institut de Physique Th\'eorique, CEA, IPhT, CNRS, URA 2306,
  F-91191Gif/Yvette Cedex, France \\ E-mail:
  \email{philippe.brax@cea.fr}}

\date{today}

\abstract{These lecture notes cover some of the theoretical topics associated with cosmic acceleration. Plausible explanations to cosmic acceleration include dark energy, modified gravity
and a violation of the Copernican principle. Each of these possibilities are briefly described.
}

\begin{document}

\section{Introduction}
More than ten years after its initial discovery\cite{Perlmutter:1998np,Riess:1998cb}, cosmic acceleration remains an unsolved problem. In fact, this phenomenon is
so much at odds with conventional particle physics and cosmology that a solution might require a complete reformulation
of the laws of physics governing both  very small scales and  very large scales. Indeed as we will see, the contemporary
models trying to explain  cosmic acceleration using quantum field theory and general relativity fail to provide a convincing framework.
In these lecture notes, I will not attempt to present the experimental and observational status of cosmic acceleration. They have been covered by
Jean-Christophe Hamilton in his lectures. I will only try to provide some indications about cosmic acceleration. The number of important topics has grown enormously in the last ten years, so much that I will
only be able to cover a limited number of them.

Cosmic acceleration was first observed using type Ia supernovae and their Hubble diagram, i.e.  the redshift vs luminosity distance. The result is purely kinematical and stipulate that the acceleration parameter of type Ia supernovae $q=-\frac{a\ddot a}{a^2}$ is negative in the recent past of the universe. This implies that distances measured according to the Friedmann-Robertson-Walker metric (FRW) are increasing fast $\ddot a >0$:
\begin{equation}
ds^2=-dt^2 + a^2(t) \left [ \frac{dr^2}{1-kr^2} + r^2 (d\theta^2 + \sin^2 \theta d\phi^2)\right ]
\end{equation}
Notice that $a$ has a dimension here and
the parameter $k=0,\pm 1$ is the reduced curvature. Spatial sections are open, closed or flat depending on $k=-1,1$ or $k=0$.
Another period of accelerated expansion seems to have also existed in the early universe and bears the name of cosmic inflation\cite{Lyth:1998xn}. Here the late time acceleration
started when $z\sim 1$ where the redshift is defined by
\begin{equation}
1+z= \frac{a_0}{a}
\end{equation}
and a subscript 0 denotes the present value of a cosmological quantity.

Understanding the observation of cosmic acceleration requires a theoretical framework. In the last century, cosmology\cite{Trodden:2004st} has been very successful in describing
the evolution of the universe using two fundamental assumptions. The first one is:\\ \\
\centerline{\it The Universe can be described using the general theory of relativity.}\\ \\
General relativity\cite{Will:2005va} has been tested in the solar system and beyond, notably in extreme astrophysical situations such as binary pulsars. So far, there is no reason to doubt its validity up to cosmological scales.
Another principle is usually necessary to simplify the analysis of the universe as a whole (the cosmological principle):\\ \\
\centerline{\it The universe is homogeneous and isotropic on large scales. }\\ \\
Of course, the universe is lumpy on small scales. Nevertheless, the appearance of small scale structures can be understood as resulting from the
growth of initial inhomogeneities. The cosmological principle is also a very robust hypothesis.

General relativity relates the energy content of the universe to its geometry. Observations have revealed four different types of energy in the universe.
Ordinary matter  is  described by the standard model of particle physics (both baryons and leptons) and is responsible for the existence of stars and galaxies. Radiation in the form of photons is the best probe we have to observe cosmic phenomena. Neutrinos are elusive particles which participate in the radioactive phenomena leading to the creation of the elements (Big Bang Nucleosynthesis). Finally, a host of phenomena including the rotation curves of galaxies seem to lead to  the existence of exotic particles in the form of dark matter. These four types of energy are enough to describe most of the history of the universe. Unfortunately, they cannot account for a period of cosmic acceleration.
The dynamics of the scale factor are governed by
\begin{equation}
\frac{\ddot a}{a}= -\frac{4\pi G_N}{3} (\rho +3p)
\end{equation}
where $G_N$ is Newton's constant, $\rho=\sum_{i=1}^4 \rho_i$ the total energy density of the four types of energy and $p=\sum_{i=1}^4 p_i$ the total pressure. Each fluid has an equation of state $w_i$ such that $p_i=w_i \rho_i$. Ordinary matter and cold dark matter have a vanishing equation of state $w_i=0$ while radiation and neutrinos have $w_i=1/3$. Of course, this implies that $\ddot a <0$.

A negative pressure is not enough to guarantee acceleration. One must impose that the total equation of state $p=w\rho$ satisfies $w<-1/3$. It happens that a fluid with this property was introduced by Einstein in order to guarantee the existence of a static and spherical universe comprising only ordinary matter.
Indeed the dynamics of the universe can be equivalently described using the Friedmann equation
\begin{equation}
H^2= \frac{8\pi G_N}{3} \rho - \frac{k}{a^2}
\end{equation}
where $H=\frac{\dot a }{a}$ is the Hubble rate.
Imagine that there exists another  fluid with $p=-\rho=\frac{\Lambda}{8\pi G_N}$ on top of ordinary matter. Then the $k=1$ Friedmann equation yields:
\begin{equation}
H^2= \frac{8\pi G_N}{3} \rho_m +\frac{\Lambda}{3} - \frac{1}{a^2}
\end{equation}
where $\rho_m$ is the matter density in the universe,
together with
\begin{equation}
\frac{\ddot a}{a}= -\frac{4\pi G_N}{3} \rho_m +\frac{\Lambda}{3}
\end{equation}
A static solution exists with $a= \frac{1}{\sqrt \Lambda}$. Unfortunately, the discovery by Hubble that the universe is expanding rules this model out.

A lesson can be learnt from Einstein's static universe. Indeed, the existence of a cosmological constant
is a key ingredient in order to obtain a repulsive behaviour in general relativity. For Einstein's universe, the negative pressure due to the cosmological constant counterbalances the gravitational attraction. If this is not the case, then the repulsive nature of the cosmological constant will lead to the acceleration of the universe. This is the simplest explanation of cosmic acceleration.

The acceleration of the universe can be easily formulated in terms of the energy content of the universe. Acceleration requires that the cosmological constant must play a dominant role. Indeed this conclusion  follows from the type Ia measurements and the WMAP results on the Cosmic Microwave Background (CMB)\cite{2008arXiv0803.0547K}. Let us denote by
\begin{equation}
\Omega_i= \frac{\rho_i}{\rho_c}
\end{equation}
where the critical energy density is
\begin{equation}
\rho_c= \frac{3H^2}{8\pi G_N}
\end{equation}
We have defined  the energy density\footnote{We will freely use the equivalence $8\pi G_N\equiv \kappa_4^2 \equiv m_{\rm Pl}^{-2}$ where
$m_{\rm Pl}\approx 2.10^{18}$ GeV is the reduced Planck mass.}  $\rho_\Lambda= \frac{\Lambda}{8\pi G_N}$, and we consider that there are now five fluids including the one corresponding to the cosmological constant.
The Friedmann equation can be rewritten
\begin{equation}
\Omega -1 = \frac{k}{a^2H^2}
\end{equation}
where $\Omega=\sum_{i=1}^5 \Omega_i$.
Observations of the CMB tells us that the universe is almost spatially flat $\Omega=1$. This follows from the location of the acoustic peaks in the CMB spectrum. Using general relativity and the cosmological principle, one can relate
the acceleration parameter to the fraction of cosmological constant $\Omega_\Lambda$ and matter $\Omega_m$ (we neglect radiation and the neutrinos)
\begin{equation}
q= \frac{\Omega_m}{2} - \Omega_\Lambda
\end{equation}
With these two relations one deduces that $\Omega_\Lambda$ does not vanish and even dominates now
\begin{equation}
\Omega_m \sim 0.3\ \ \Omega_\Lambda \sim 0.7
\end{equation}
 The universe is accelerating now, and  the energy content of the universe is dominated by a pure cosmological
constant. This result which was first deduced about 1998 has been greatly refined since then using CMB data, large scale structures and
type Ia supernovae. Part of these lectures will be devoted to understanding the physics behind the cosmological constant.

Observations give that
\begin{equation}
\rho_\Lambda \approx 10^{-48}\ ({\rm GeV})^4
\end{equation}
This density is approximately $10^{-29} {\rm g/cm^3}$. It is an extremely small scale which is only matched by the neutrino mass $m_\nu \sim 10^{-3}$ eV. It is extremely difficult to justify the existence of such a small cosmological constant using quantum field theory alone\cite{Weinberg:1988cp}: this is {\bf the cosmological constant problem}.
Postulating the existence of a small cosmological constant is so problematic that other possibilities have been considered. They fall within two categories. The first one amounts to introducing a new type of matter called dark energy whose role mimics a pure cosmological constant. The Einstein equations governing the evolution of the universe
\begin{equation}
R_{\mu\nu}-\frac{1}{2} g_{\mu\nu} R= 8\pi G_N T_{\mu\nu}
\end{equation}
where $R_{\mu\nu}$ is the Ricci tensor, $R$ the Ricci scalar and $T_{\mu\nu}$ the energy-momentum tensor of the four types
of energy, is supplemented with a new energy-momentum tensor due to the dark energy component of the universe
\begin{equation}
R_{\mu\nu}-\frac{1}{2} g_{\mu\nu} R= 8\pi G_N (T_{\mu\nu}+T_{\mu\nu}^{\rm DE})
\end{equation}
Several examples of dark energy will be presented.
Another possibility concerns gravity itself. Indeed, one could envisage that gravity which has been tested up to galactic scales must be modified on larger scales. This could be formulated with a modified Einstein equation
\begin{equation}
f(R,R_{\mu\nu}, R_{\alpha\beta\rho\sigma})_{\mu\nu}=8\pi G_N T_{\mu\nu}
\end{equation}
where $f_{\mu\nu}$ is a tensor involving all the components of the Ricci tensor $R_{\alpha\beta\rho\sigma}$
which reduces to the Einstein tensor $G_{\mu\nu}=R_{\mu\nu}-\frac{1}{2} g_{\mu\nu} R$  on small scales (up to the galactic scales). As we will see, this approach is fraught with difficulties.

These models are easier to describe using
the Lagrangian formulation of general relativity
\begin{equation}
S= \frac{1}{16\pi G_N}\int {\rm d}^4 x\ \sqrt{-g} (R-2\Lambda)
\end{equation}
The first term is the Einstein-Hilbert action. The second term involves the cosmological constant. Notice that it is a simple additive correction to the Einstein-Hilbert action. In fact, this additive constant plays the same role as a vacuum energy.
This degeneracy is the origin of the cosmological constant problem.

Using the action principle, it is easy to understand how to incorporate a dark energy component, one considers the action
\begin{equation}
S= \frac{1}{16\pi G_N}\int {\rm d}^4 x\ \sqrt{-g} (R-{\cal L}_{\rm DE})
\end{equation}
where ${\cal L}_{\rm DE}$ is the dark energy Lagrangian which reduces to a constant on cosmological scales. We will give many examples of dark energy
Lagrangians.
Modifying gravity can be easily implemented too, it only amounts to altering the Einstein-Hilbert term
\begin{equation}
S= \frac{1}{16\pi G_N}\int {\rm d}^4 x\ \sqrt{-g} h(R,R_{\mu\nu}, R_{\mu\nu\rho\sigma})
\end{equation}
where $h$ is a scalar function. Different forms of $h$ will be presented too.

So far, we have assumed that the Einstein equations need to be modified in order to accommodate the acceleration of the universe. In fact, there is another appealing explanation which only requires to assume that the cosmological principle is violated. In particular, if we lived in a large void, the acceleration of the universe would be only apparent and due to matter depletion in the void. Of course, this would imply that observers have a special place in the universe, contrary to the Copernican principle. In the course of these lectures, we will encounter other even more drastic routes which have been followed in order to understand cosmic acceleration; for instance we will discuss
DGP gravity\cite{Dvali:2000hr}  and unimodular gravity\cite{Shaposhnikov:2008xb}.

The acceleration of the universe is a rich subject and I cannot cover all the diverse solutions which have been proposed. The following choice of topics is personal. More thorough coverage of the subject can be found in excellent review articles\cite{Caldwell:2009ix,Silvestri:2009hh,Copeland:2006wr}. Citations will unfortunately be parsimonious, mainly general references where more information can be found.

In a first part I shall recall  the status of the cosmological constant problem. I shall then describe dark energy and its siblings the scalar-tensor theories including $f(R)$ gravity. Another chapter is devoted to modified gravity in the infrared. Finally, the possibility of a violation of the Copernican principle will be analysed briefly.

\section{ The Cosmological Constant}

Cosmic acceleration may be explained by the presence of a cosmological constant supplementing Einstein's general relativity. From the observational point of view, this is by far the simplest and most economic explanation. From a fundamental point of view, the cosmological constant term is puzzling. As mentioned, a cosmological constant plays the same role as a vacuum energy. Quantum theory has taught us that the vacuum is not empty space: it is full of vacuum fluctuations in the guise of particle-antiparticle creations over extremely short time scales. These fluctuations have had a nice experimental confirmation in the Casimir effect whereby two metallic plates attract each other under the influence of electromagnetic quantum effects. As a result, one should think of the cosmological constant as an energy density comprising two terms
\begin{equation}
\rho_{\Lambda}=  {\Lambda_0}m_{\rm Pl}^2 + V_0
\end{equation}
where $V_0$ contains all the quantum fluctuations due to the particle physics vacuum and
the term ${\Lambda_0}m_{\rm Pl}^2$ is called the bare cosmological constant.
The net result given by observations is that $\rho_{\Lambda}$ is very small. How can this be?

\subsection{The cosmological constant problem}

Vacuum contributions to the cosmological constant arise from the creation and annihilation of particle-antiparticle pairs. Each particle $i$ of mass $m_i$ contributes
\begin{equation}
V_0= \sum_{i} \Delta V_{i}, \ \ \  \ \Delta V_i= \frac{c_i}{16\pi^2} m_i^4
\end{equation}
where $c_i=O(1)$.
Taking into account known particles such as the gauge bosons $Z$ and $W$, one gets a contribution of order $M_Z^4$ which is sixty orders of magnitude larger than the observed $\rho_{\Lambda}$.
Facing such a major difficulty, two logical possibilities can be envisaged. First of all, there could be a cancelation effect between $V_0$ and
${\Lambda_0}m_{\rm Pl}^2$. This would require a very precise fine tuning of sixty orders of magnitude. Another possibility is that both $V_0$ and
${\Lambda_0}m_{\rm Pl}^2$ are small.

Large cancelations are highly unnatural in quantum field theory. For this reason, an almost exact cancelation between $V_0$ and ${\Lambda_0}m_{\rm Pl}^2$ must have another origin. A clue is given by Weinberg's bound on $\rho_{\Lambda}$. Suppose that observations did not indicate that $\rho_\Lambda \approx 10^{-48} ({\rm GeV})^4$. Could we infer a reasonable value for $\rho_{\Lambda}$? This is the question which was positively answered by Weinberg\cite{Weinberg:1988cp}. Indeed as soon as a cosmological constant term starts dominating the dynamics of the universe, structures stop growing. This implies that galaxies must have formed before the start of an accelerated period. This leads to a bound
\begin{equation}
\rho_{\Lambda} \le 500 \rho_{m0}
\end{equation}
where $\rho_{m0}$ is the present matter density. In fact this entails that $\rho_{\Lambda}\le 10^{-46} {\rm GeV}^4$, i.e. a very stringent bound.
Thus a strong bound on the cosmological constant results from the existence of galaxies only. This reasoning can be extended and stated as
an anthropic result: the existence of observers is only compatible with a small cosmological constant. In a sense, the mere fact that we observe
the universe imposes that the cosmological constant must be small. As we will see in the next section, this interpretation can be put on firmer ground
when a large number of possible vacua exist, and if the probability of existence of these vacua is evenly distributed, one could argue that there is
a non-negligible probability that our universe is atypical and happens to accommodate a large cancelation between quantum effects and the bare cosmological constant.

Within the realm of quantum field theory, the previous cancelation is unnatural. To this end, it is useful to recall a few fact about effective field theories\cite{Burgess:2004ib,Burgess:2005wu}. Let us  consider a quantum field theory describing high energy phenomena up to a  scale $E$. In this high energy theory, a bare cosmological constant is present $\rho_\Lambda(E)$. Let us now consider the physics at  lower scales $\mu<E$. When lower energies are probed, particles of intermediate masses
$\mu\le m_i\le E$ cannot be produced in experiments. They can only appear in the vacuum fluctuations. Hence at the lower energy scale $\mu$, the physics can be described by an effective field theory comprising only particles whose masses are $m_i\le \mu$ and a cosmological constant term
\begin{equation}
\rho_{\Lambda}(\mu)= \rho_\Lambda (E)+ \sum_{\mu\le m_i \le E} \frac{c_i}{16\pi^2} m_i^4
\end{equation}
In this context, the bare cosmological constant $\Lambda_0=\kappa_4^2 \rho_\Lambda(E)$ is a parameter depending on high energy physics whereas the
quantum fluctuations involve particles which have been integrated out. Describing the acceleration of the universe requires to consider low energies well below the electron mass $\mu\le m_e$.
Viewed from this angle, an exact cancelation between both terms  is hard to justify and would require very special properties.
It is more natural to assume that both $\rho_\Lambda(E)$ and the quantum corrections are small. We will see that this is also extremely difficult to realise.

\subsection{The Landscape}

Recently, anthropic ideas and string theory\cite{Bousso:2000xa} have led to the landscape picture whereby a large number of vacua could explain the existence of
a small cosmological constant. Although the string picture is too complex to present here, we can extract some of its key ingredients
in a 4d context and a conventional field theory.
Let us consider the action
\begin{equation}
S= \int {\rm d}^4 x \sqrt{-g} (\frac{R}{2\kappa_4^2} -\rho_{\Lambda}(m_e)- \frac{1}{4!} \sum_{i=1}^J F_i^2)
\end{equation}
where we have taken into account $\rho_{\Lambda}(m_e)$ at the electron mass for the sake of the argument. The action involves a collection of $J$ four forms $F^{\mu\nu\rho\sigma}_i$.
The equations of motion for the four forms yield $F^{\mu\nu\rho\sigma}_i= n_i q \epsilon^{\mu\nu\rho\sigma}$ where $n_i$ is an integer (this is analogous to the Dirac charge quantisation). This immediately leads to an effective vacuum energy
\begin{equation}
\rho_{\Lambda}= \rho_{\Lambda}(m_e) -\frac{q^2}{2} \sum_{i=1}^J n_i^2
\end{equation}
Each choice of the integers $n_i$ correspond to a different vacuum. In fact, if we consider the state vector $ \{ n_1\dots n_J \}$ as representing
one type of universe with its particular cosmological constant, can we arrange the sum $\frac{q^2}{2} \sum_{i=1}^J n_i^2$ to cancel $\rho_{\Lambda}(m_e)$? This would select  the type of universe we live in.

The number $N$ of vacua in the interval $[n^2,n^2+ dn^2]$ with  $n^2= \sum_{i=1}^J n_i^2$ is related to  the area of a J-sphere of radius $n$, i.e.
\begin{equation}
\frac{dN}{dn^2}= \frac{(2\pi)^{J/2}}{2\Gamma(J/2)} n^{J-2}
\end{equation}
Increasing $N$ by one unit requires $\delta n^2= \frac{dn^2}{dN}$, this is the typical spacing between values of $n^2$ in the space of all vacua.
This corresponds to a step in the cosmological constant
\begin{equation}
\delta \rho_{\Lambda}= - \frac{q^2}{2} \frac{ 2 \Gamma (J/2)}{ (2\pi)^{J/2}} n^{2-J}
\end{equation}
Typically, one would like to cancel $\rho_{\Lambda}(m_e)$ which  requires to cancel a large number
\begin{equation}
n^2=\frac{2 \rho_{\Lambda}}{q^2}
\end{equation}
This cancellation cannot be exact as the right-hand side is not always an integer. This is almost exact if the interval between each
step is very small, what is left over being interpreted as the energy density driving cosmic acceleration.
The step size is simply
\begin{equation}
\frac{\delta \rho_{\Lambda}}{\rho_{\Lambda}(m_e)}=  \frac{ 2 \Gamma (J/2)}{ (2\pi)^{J/2}} n^{-J}
\end{equation}
This ratio measures the amount of fine tuning in order to cancel the cosmological constant approximately and gives a bound on the leftover cosmological constant which drives the cosmic acceleration. It decreases with an increasing number of forms $J$. Canceling sixty orders of magnitude requires
$\frac{\delta \rho_{\Lambda}}{\rho_{\Lambda}(m_e)}=10^{-60}$ which can easily be achieved with a large number of forms.

This toy model is a good illustration of the landscape paradigm. Indeed a large number of four forms can almost cancel the cosmological constant, leaving a tiny remainder whose presence leads to the current cosmic acceleration. Although the adjustment is fine tuned, there is a non-vanishing number of vacua specified by the integers $n_i$ which can satisfy these constraints. If all the possible vacua can be populated, the anthropic principle specifies that we simply happen to live in one of these particular universes.
\subsection{Weinberg's theorem}

So far we have not tried to have a dynamical understanding of cosmic acceleration. Indeed the landscape picture just postulates that the overall effect of the bare cosmological constant and quantum fluctuations can be almost completely canceled by negative contributions due to many four forms.
From the point of an effective field theory, this is not satisfying and appears as a last resort explanation. A more natural explanation would be the existence of a symmetry principle which would guarantee that the cosmological constant vanishes. This symmetry may be slightly broken resulting in a tiny cosmological constant whose value would drive the cosmic acceleration. This is the technical meaning of natural in field theory. A theory is natural if by taking a parameter to zero in a Lagrangian, the degree of symmetry of the theory increases. We will present two such symmetries and analyse their drawbacks: scale invariance\cite{Weinberg:1988cp,Burgess:2004ib,Burgess:2005wu} and supersymmetry.

The cosmological constant $\rho_{\Lambda}$ has mass dimension four. This implies that a change of scale $x\to \lambda x$ rescales
$\rho_{\Lambda} \to \lambda^{-4} \rho_{\Lambda}$. Of course, if the field theory describing  cosmic acceleration is scale invariant, the resulting cosmological constant must vanish
\begin{equation}
\rho_{\Lambda}\equiv 0
\end{equation}
The simplest field theories which could model out cosmic acceleration are scalar field theories. Indeed scalar fields $\phi_i$ do not carry Lorentz indices and can then acquire a vacuum expectation value (vev) without breaking Lorentz invariance. Scalar field theories can be specified by their interaction potential $V(\phi_i)$. In particular we always include the bare cosmological constant as a constant term in the  potential. Vacua of these theories minimise the potential
\begin{equation}
\partial_i V\vert_{\phi_i=<\phi_i>}=0
\end{equation}
We assume that the potential $V(\phi_i)$ is the low energy effective potential valid for scales well below the electron mass. Hence it captures the effect of integrating out massive particles. The vev of the effective potential is then interpreted as the effective cosmological constant
\begin{equation}
\rho_{\Lambda}= V(<\phi_i>)
\end{equation}
Now let us assume that the low energy effective action is classically scale invariant. This implies that the potential can be expanded as
\begin{equation}
V(\phi_i)= \lambda_{ijkl}\phi_i\phi_j\phi_k \phi_l
\end{equation}
in terms of monomials of dimension 4. Each scalar field has dimension one.
We can always single out one field $\phi_1$ and write
\begin{equation}
V(\phi_i)= \phi_1^4 \tilde V(z_i)
\end{equation}
where $z_i=\phi_i/\phi_1$ and $\tilde V(z_i) = V(1,z_2\dots z_N)$.
Vacua of this theory satisfy
\begin{equation}
\partial_{z_i} \tilde V =0, \ \tilde V=0,
\end{equation}
i.e. they have a vanishing cosmological constant. The potential $\tilde V$ is a function of $(N-1)$ variables. The existence of a minimum
with a vanishing cosmological constant is only possible if there exists a relation amongst the coupling constants
\begin{equation}
g(\lambda_{ijkl})=0
\end{equation}
Notice too that the vev of $\phi_1$ is not specified, it is a Goldstone direction corresponding to the spontaneous breaking of scale invariance by the non-trivial vacuum $<z_i>$. The origin where all the fields $\phi_i$ vanish is also a vacuum of the theory. In fact, writing $<\phi_i>= \ \phi_1 <z_i>$, one can see that there is a flat direction of vacua in field space parameterised by $\phi_1$. Along this flat direction, the cosmological constant vanishes identically.

The scalar field $\phi_i$ can be used to give masses to fermions thanks to Yukawa couplings
\begin{equation}
{\cal L}_{\rm yuk}= y_{ijk} \phi_i \bar \psi_j \psi_k
\end{equation}
leading to fermion masses $m_{jk}= y_{ijk}<\phi_i>$. Apart from the origin, all the vacua along the flat direction lead to non-vanishing masses.
Although the effective action at low energy is obtained by integrating out massive particles and only contains massless fields before spontaneous breaking of scale invariance, quantum corrections are still present and lead to a logarithmic running of the couplings constant $\lambda_{ijkl}$.
This scale dependence implies that generically
\begin{equation}
g(\lambda_{ijkl}(\mu))\ne 0
\end{equation}
unless another symmetry is present and guarantees that the relation between the coupling constants is invariant under the renormalisation group
evolution. If this is not the case, the vacuum $<z_i>$ is lifted by quantum corrections. Of course, there always remains the trivial vacuum where all the fields vanish. Unfortunately this vacuum cannot give masses to fermions and is therefore not suitable to describe our universe.

Weinberg's theorem is very powerful as it rules out most attempts to find a stable vacuum with a vanishing cosmological constant. On the other hand, it also specifies that under special circumstances, such vacua may exist. A particularly illuminating example is provided by supersymmetry.

\subsection{Supersymmetry}

Supersymmetry\cite{Martin:1997ns} has been used  to solve the hierarchy problem in particle physics, i.e. it stabilises the mass of the Higgs boson under quantum corrections. Of course, with a certain twist of fate, the same type of argument can be applied to show that a vanishing cosmological constant is stable under quantum corrections in supersymmetric models.
Supersymmetric models postulate that each particle has a superpartner called a sparticle. Fermions of spin $1/2$ have scalar partners while gauge fields have fermionic partners.
The supersymmetry algebra is simply given by the anticommutation relation
\begin{equation}
\{ Q_\alpha, \bar Q_{\dot\alpha}\}= 2\sigma^\mu_{\alpha\dot\alpha} P_\mu
\end{equation}
where $Q_\alpha$ are fermionic generators and their complex conjugates $ Q_{\dot\alpha}$ (they are Weyl spinors with $\alpha=1,2$), $P_\mu$ is the momentum operator and $\sigma^\mu$ are the three Pauli matrices $\sigma^i$ and $\sigma^0=1$.
The vacuum is represented by a state vector $\vert 0>$ which is invariant under supersymmetry, i.e. it is annihilated by
\begin{equation}
Q_\alpha \vert 0>=0
\end{equation}
Specifying $\mu=0$ and taking the vacuum expectation value of the anticommutation relation
leads to
\begin{equation}
<0\vert P_0 \vert 0>=0
\end{equation}
where we have used the fact that the vacuum is translation invariant $P_i\vert 0>=0$. Of course $P_0=H$ is the energy operator (the Hamiltonian), and therefore we conclude that
\begin{equation}
\rho_{\Lambda}\equiv 0
\end{equation}
Strong non-renormalisation theorems exist in supersymmetry which guarantee that this is true to all order in perturbation theory. In a nutshell this follows from the existence of as many bosons and fermions of equal masses. Loop corrections due to bosons are exactly canceled by their fermionic
counterparts.

As a result supersymmetric theories are strongly motivated candidates to explain cosmic acceleration. Let us briefly summarise how they can be built. We focus on the scalar sector only. The theories are specified by two functions  $W(\phi^i)$ where the scalar fields are complex and
$K(\phi^i,\bar \phi ^{\bar i})$. The superpotential $W$ and the K\"ahler potential $K$ can be combined to give the scalar potential
\begin{equation}
V_F= K^{i\bar j} F_i \bar F_{\bar j}
\end{equation}
where $K^{i\bar j}$ is the inverse matrix of $K_{i\bar j}= \partial_i \partial_{\bar j} K$ and
\begin{equation}
F_i=\partial_i W
\end{equation}
To simplify the discussion, we  assume that all the scalar fields are neutral.
The kinetic terms are simply
\begin{equation}
{\cal L}_{\rm kin}= K_{i\bar j} \partial_\mu \phi^i \partial^\mu \bar\phi^{\bar j}
\end{equation}
Notice that the potential is always positive $V_F\ge 0$ and that supersymmetric vacua are obtained when $F_i=0$. This implies that the cosmological constant vanishes.

Unfortunately supersymmetry is not a symmetry of nature. Indeed no superpartner has ever been observed implying that there must be a mass splitting between superpartners. This can be achieved when supersymmetry is broken. Denoting by $M_{\rm SUSY}$ the supersymmetry breaking scale, the discrepancy between the superpartners leads to a contribution to the cosmological constant
\begin{equation}
\delta \rho_{\Lambda}\vert_{\rm SUSY} \sim M^2_{\rm SUSY}E^2
\end{equation}
where $E$ is the largest energy scale described by the supersymmetric model. As $E\ge M_{\rm SUSY}\ge M_Z$, we find that the breaking of supersymmetry reintroduces a serious fine tuning issue.

Moreover, supersymmetry must be supplemented with a gravitational sector in order to incorporate the Einstein-Hilbert action of general relativity. In this case, the potential becomes
\begin{equation}
V_F= K^{i\bar j} F_i \bar F_{\bar j} -3 m_{3/2}^2 m_{\rm Pl}^2
\end{equation}
where
\begin{equation}
F_i= e^{\kappa_4^2 K/2} (\partial_i W + \kappa_4^2 (\partial_i K) W)
\end{equation}
and  the gravitino mass is
\begin{equation}
m_{3/2}=e^{\kappa_4^2 K/2}\vert  W\vert
\end{equation}
The gravitino is the spin $3/2$ superpartner of the graviton.
In supergravity the potential is not positive anymore. Moreover, supersymmetric vacua still satisfy $F_i=0$ and have a negative cosmological constant
\begin{equation}
\rho_{\Lambda}= -3 <m_{3/2}^2> m_{\rm Pl}^2
\end{equation}
leading to an anti de Sitter geometry. Broken supergravity can have a vanishing cosmological constant provided one fine-tunes the gravitino mass to balance the $F$-term contribution.

\section{Dark Energy}
We have  just seen that symmetry arguments are not enough to guarantee the existence of a vacuum with an almost vanishing cosmological constant. It could well be that the universe is not in a vacuum state at all and has  a dynamical evolution. In this case the energy density responsible for the acceleration of the universe need not be constant. This evolving energy density has been called dark energy.

\subsection{ A fluid approach}

The cosmological constant can be viewed as a fluid pervading the entire universe and possessing an equation of state
\begin{equation}
w=\frac{p}{\rho}
\end{equation}
which is constant and equal to -1. In fact, cosmic acceleration does not require the equation of state to be $-1$. We shall first consider
fluids for which the equation of state $w$ is constant. Accelerating happens as soon as
\begin{equation}
w\le -\frac{1}{3}
\end{equation}
A natural restriction on the possible value of the equation of states arises from the weak energy condition in general relativity:
$T^{\mu\nu}t_\mu t_\nu\ge 0$ for any time-like vector $t^\mu$. In other words, the energy momentum $T^{\mu\nu}$ must be positive along any
particle trajectory described by a time-like tangential vector $t^\mu$. The weak energy condition implies that
\begin{equation}
\rho\ge 0, \ w\ge -1
\end{equation}
where the energy momentum tensor of a perfect fluid is
\begin{equation}
T_{\mu\nu}= (\rho+p) u_\mu u_\nu + pg_{\mu\nu}
\end{equation}
and $u_\mu$ is the velocity four-vector of the fluid.
Conservation of matter $D_\mu T^{\mu\nu}=0$ provides the equation
\begin{equation}
\dot \rho=-3H(\rho+p)
\end{equation}
Solving the conservation equation gives the evolution
\begin{equation}
\rho= (\frac{a}{a_0})^{-3(1+w)}\rho_0
\end{equation}
Using the Friedmann equation for a flat model, one finds
\begin{equation}
a=(\frac{t}{t_0})^{2/3(1+w)}a_0
\end{equation}
Of course, this is only valid when dark energy dominates.

Nothing prevents the equation of state to depend on the energy density too $w(\rho)$. These phenomenological  models lead to a time-evolution of the equation of state. The simplest of these models is the Chaplygin gas\cite{Kamenshchik:2001cp} with
\begin{equation}
p=-\frac{A^2}{\rho}
\end{equation}
The conservation equation can be solved and gives
\begin{equation}
\rho=\sqrt{A^2+ \frac{B^2}{a^6}}
\end{equation}
Early in the universe when $a$ is small $\rho \sim B/a^3$ like for cold dark matter, while at late time $\rho\sim A$, i.e. a pure cosmological constant. The emergence of the cosmological constant at late time  could explain why acceleration is only a very recent phenomenon. Unfortunately, this model is under pressure due to its poor properties in explaining structure formation. Generalisations with $\rho=-A^2/\rho^\alpha$ are acceptable provided $\alpha\le 0.2$\cite{Bean:2003ae}.

Models with an equation of state $w<-1$ violate the weak energy condition and possess ghosts. This leads to a catastrophic instability as the scale factor behaves like
\begin{equation}
a(t)= (1-\frac{t}{t_{\rm end}})^{2/3(1+w)}
\end{equation}
The universe reaches a singularity in a finite time  $t_{\rm end}$ where the Hubble rate diverges: the big rip\cite{Caldwell:2003vq}.

\subsection{Scalar field models}

The fluid approach to dark energy is phenomenological. A more fundamental description which could be unified with the standard model of particle physics can be obtained using a quantum field theory approach.
Assuming that Lorentz invariance is not broken in the quantum field theory describing dark energy, the fact that the dark energy density has a time dependence implies that it must have a spatial dependence too. The simplest models for such a space-time dependent energy density are scalar field theories. In the following, we shall concentrate  on scalar models of dark energy.

For simplicity sake, we focus on single field models
\begin{equation}
S=\int {\rm d}^4 x \sqrt{-g}[\frac{R}{2\kappa_4^2} - \partial_\mu\phi\partial^\mu \phi -V(\phi)]
\end{equation}
The equation of motion for the scalar field in a FRW background is the Klein-Gordon equation
\begin{equation}
\ddot\phi+ 3H\dot\phi +\frac{dV}{d\phi}=0
\end{equation}
Notice that the expansion of the universe acts as a friction term: the Hubble friction. Single scalar field models can be rephrased into a fluid
picture where the energy density and the pressure are given by
\begin{equation}
\rho_\phi=\frac{\dot \phi^2}{2} + V(\phi),\ \ p_\phi= \frac{\dot \phi^2}{2} -V(\phi)
\end{equation}
Notice that the equation of state is always $-1\le w_\phi\le 1$.
The dynamics of dark energy can be nicely classified in two categories: thawing and freezing models\cite{Caldwell:2009ix}.

Thawing models are such that the field starts
at a non-zero value of the potential and stays there due to the Hubble friction. When the Hubble friction is small enough, the field starts moving and rolls down towards lower values of the potential. Several possibilities can be envisaged. If the potential possesses minima, the field will converge towards one of these minima. Depending on the sign of the potential at the minimum where the field asymptotically settles down, the universe becomes
either a de Sitter space with an eternally accelerated evolution if $V_{\rm min} >0$, or flat space with no cosmological constant if $V_{\rm min}=0$.
If the minimum has  negative energy, the universe ends up in a big crunch singularity. Phenomenologically potentials leading to $V_{\rm min}=0$ are favoured.
Freezing models are such that the potential has no minimum. The field keeps rolling down the potential slope in a decelerating fashion until it virtually stops under the effect of the Hubble friction.

In both cases, the mass of the scalar field needs to be related to the Hubble rate. Indeed, if the mass of the scalar field identified with
\begin{equation}
m^2_\phi= \frac{d^2 V}{d\phi^2}
\end{equation}
is large compared to $H$, the field rolls very fast along the potential. The kinetic terms dominate and the equation of state is close to $w_\phi \approx 1$. On the contrary, if the mass is very small compared to the Hubble rate, the field is in a slow roll state. The kinetic energy is negligible and $w_\phi \approx -1$, i.e. the model mimics a pure cosmological constant. The only possibility to obtain a dark energy behaviour with an evolution of the energy density is to satisfy
\begin{equation}
m_\phi\approx H
\end{equation}
This is a strong constraint on the type of dark energy models as it states that the mass of the scalar field now must be tiny $m_\phi\approx  H_0\sim 10^{-43}$ GeV.

\subsection{ Attracting and tracking }

In this section we will describe two families of very useful potentials leading to dark energy of the freezing type. Thawing models will be exemplified later.
The first family consists of inverse power law potentials\cite{Ratra:1987rm}
\begin{equation}
V= \frac{M^{4+n}}{\phi^n}
\end{equation}
The behaviour of dark energy with such a Ratra-Peebles potential is interesting as the long time behaviour of the scalar field
is largely independent of the initial conditions. Indeed, let us consider the radiation or the matter eras where the dominant fluid
has an equation of state $w_B=1/3$ or $w_B=0$. In this regime, the Klein-Gordon equation has an exact solution
\begin{equation}
\phi\sim t^{-\frac{2}{n+2}}
\end{equation}
corresponding to an energy density
\begin{equation}
\rho_\phi\sim t^{-\frac{2n}{n+2}}
\end{equation}
Moreover this solution is an attractor. Choosing any initial solution for $\phi$ in the early universe, the solution always converges to this attracting solution.

During this regime the energy density decays as
\begin{equation}
\rho_\phi\sim a^{-3\frac{n(1+w_B)}{2+n}}
\end{equation}
This corresponds to an equation of state
\begin{equation}
w_\phi= \frac{nw_B-2}{n+2}
\end{equation}
 Notice that during the matter era, the energy density decays like
$a^{-3n/(n+2)}$ which is slower than the matter evolution in $a^{-3}$. This implies that the dark energy density eventually catches up with the
matter density and becomes dominant. When this is the case, the solution of the Klein-Gordon equation leaves the attractor (only valid when matter or radiation dominate) and the universe starts accelerating. One of the nice features of this model is the independence of initial conditions. Dark energy becomes always dominant at some point.

The value of the field now is also easily determined noticing that the mass $m_\phi\sim V(\phi)/\phi^2$. Using the fact that $V(\phi_{0}) \sim \rho_{c0}$ and $m_\phi\sim H_0$, we find that
\begin{equation}
\phi_0\sim m_{\rm Pl}
\end{equation}
Hence independently of the initial conditions, the field freezes around a value close to the Planck scale now. Of course this can only happen if the scale $M$ in the potential is such that
\begin{equation}
M^{4+n} \approx \rho_{c0}m_{\rm Pl}^{n}
\end{equation}
Once the scale $M$ is fixed, the dynamics of the model follow: irrespective of initial conditions the field $\phi$ reaches the attractor during both the radiation and matter eras. It leaves the attractor in the recent past and stops at a value close to the Planck scale where acceleration starts.

Another interesting class of models involves exponential potentials\cite{Ferreira:1997hj,Wetterich:1987fm}
\begin{equation}
V(\phi)=M^4 \exp (-\lambda \kappa_4 \phi)
\end{equation}
where $M$ is an overall scale and $\lambda>0$. The potential decreases towards zero at infinity. There are two remarkable regimes for these models.
First of all if dark energy dominates, an attractor solution with $\Omega_\phi=1$ can be obtained provided
\begin{equation}
\lambda^2< 3(1+w_B)
\end{equation}
In this case, the asymptotic behaviour of the scalar field energy density is such that the equation of state is
\begin{equation}
w_\phi=  -1 + \frac{\lambda^2}{3}
\end{equation}
Acceleration occurs when
\begin{equation}
\lambda < \sqrt 2
\end{equation}
In these models, the present state of the universe is transient, matter  becomes more and more diluted until
eventually the energy density of the universe becomes scalar field dominated.

For larger values of $\lambda$
\begin{equation}
\lambda^2>3(1+w_B)
\end{equation}
the scalar field energy density has the same equation of state as the background fluid
\begin{equation}
w_\phi=w_B
\end{equation}
and the scalar field energy density fraction is fixed
\begin{equation}
\Omega_\phi= \frac{3(1+w_B)}{\lambda^2}
\end{equation}
i.e. the ratio of the matter energy density over the scalar field energy density is fixed. This solution is said to be scaling.
The tracking solution is such that the background fluid and the scalar field go hand in hand, hence the name tracking. Of course, in this regime, acceleration cannot occur.
This can be remedied when two exponential terms are present\cite{Copeland:2006wr}
\begin{equation}
V(\phi)=M^4(\exp (-\lambda_1 \kappa_4 (\phi-\phi_c))+ \exp (-\lambda_2 \kappa_4 \phi))
\end{equation}
Assume that $\lambda_1^2>3(1+w_B)$ and $\lambda_2^2 <2$. As long as
$\phi<\lambda_1\phi_c/(\lambda_1-\lambda_2)$ the potential is dominated by the first exponential. The field converges to the tracking behaviour and follows the background evolution. Eventually, the field overcomes $\phi>\lambda_1\phi_c/(\lambda_1-\lambda_2)$ and the second exponential kicks in leading to an accelerated expansion.
Of course these models suffer from a fine tuning issue as the scale $M$ or equivalently the initial value of $\phi$ must be tuned to reach $\Omega_\phi$ now.

Another class of models is particularly useful. The potential is given by\cite{Albrecht:1999rm}
\begin{equation}
V(\phi)= V_0 e^{-\lambda \kappa_4 \phi} ( a+ (\kappa_4\phi -b)^2)
\end{equation}
where $\lambda^2>3(1+w_B)$.
These Albrecht-Skordis potentials are such that in the early universe the exponential behaviour dominates and the scalar field tracks the background fluid. Later the field feels the presence of a local minimum and gets trapped there mimicking the role of a pure cosmological constant.

For more general potentials, there is a very useful criterion allowing one to distinguish models leading to the asymptotic domination by the scalar field and therefore cosmic acceleration\cite{Steinhardt:1999nw}. Defining $x= \frac{\kappa \dot \phi}{\sqrt 6 H}$, the effective coupling constant
\begin{equation}
\lambda_\phi=-\frac{d\ln V}{d\kappa_4\phi}
\end{equation} satisfies the
differential equation
\begin{equation}
\frac{d\lambda_\phi}{dt}= -\sqrt 6 \lambda^2 (\Gamma -1) H x
\end{equation}
Assuming that the field rolls towards large values, and provided
\begin{equation}
\Gamma >1
\end{equation}
where
\begin{equation}
\Gamma= \frac{ VV''}{V'^2}
\end{equation}
with $'=d/d\phi$, the coupling constant $\lambda_\phi$ converges towards the fixed point $\lambda_\phi=0$. In this case, the slope of the
potential vanishes,  the scalar field energy dominates eventually and acceleration occurs.
For instance for Ratra-Peebles potentials
\begin{equation}
\Gamma=1 +\frac{1}{n}>1
\end{equation}
while $\Gamma$ diverges at the minimum for Albrecht-Skordis potentials.

\subsection{Coupled dark energy}

So far dark energy has been completely decoupled from the rest of physics. This is not always the case, especially in models derived from string theory or extra dimensions. As a simplified first step, let us focus on a direct coupling between dark energy and cold dark matter\cite{Amendola:1999er}. In this case, both the dark matter energy density and the dark energy one are not separately conserved but satisfy
\begin{equation}
\dot \rho_m + 3H \rho_m = \alpha_\phi \rho_m \dot \phi
\end{equation}
and
\begin{equation}
\dot \rho_\phi + 3H(1+w_\phi) \rho_\phi = -\alpha_\phi \rho_m \dot \phi
\end{equation}
These equations will be explained when we deal with scalar-tensor theories.
We restrict our attention to single exponential models with a coupling $\lambda$. The  behaviours obtained in the decoupled case can be generalised.  The scalar field
reaches an attractor with $\Omega_\phi=1$ when
\begin{equation}
\lambda<\frac{(\alpha_\phi^2+12(1+w_B))^{1/2}- \alpha_\phi}{2}
\end{equation}
In this case the equation of state is
\begin{equation}
w_\phi= -1 +\frac{\lambda^2}{3}
\end{equation}
with acceleration for $\lambda^2<2$.
When $\lambda$ is larger
\begin{equation}
\lambda>\frac{(\alpha_\phi^2+12(1+w_B))^{1/2}- \alpha_\phi}{2}
\end{equation}
the scalar field has a scaling behaviour where the ratio of the matter and scalar field energy densities is constant. It is convenient to define an effective equation of state
\begin{equation}
w_{\rm eff}= \frac{ w_\phi \rho_\phi+ w_B \rho_B}{\rho_\phi+ \rho_B}
\end{equation}
given by
\begin{equation}
w_{\rm eff}= \frac{w_B \lambda- \alpha_\phi}{\lambda+\alpha_\phi}
\end{equation}
Acceleration can occur when
\begin{equation}
\alpha_\phi> \frac{\lambda(1+ w_B)}{2}
\end{equation}
In this case, the energy density fraction of the scalar field is
\begin{equation}
\Omega_\phi=\frac{\alpha_\phi(\alpha_\phi+\lambda)+ 3(1+w_B)}{(\alpha_\phi+\lambda)^2}
\end{equation}
For large values of $\alpha_\phi$, one can get $\Omega_\phi\sim 0.7$.
Unfortunately explicit models using this attractor suffer from phenomenological problems such as a short matter dominated era.

\subsection{Phantom dark energy}

Scalar field models can easily have an equation of state $w_\phi<-1$. Consider the action
\begin{equation}
S=\int {\rm d}^4 x \sqrt{-g}[\frac{R}{2\kappa_4^2} + \partial_\mu\phi\partial^\mu \phi -V(\phi)]
\end{equation}
where the sign of the kinetic terms has been reversed. The equation of state becomes
\begin{equation}
w_\phi= -\frac{\frac{\dot \phi^2}{2} + V(\phi)}{ \frac{\dot \phi^2}{2} -V(\phi)}
\end{equation}
which can satisfy $w_\phi \le -1$. We have already seen that this would lead to a cosmological big rip singularity. One can even show that
the effective field theory with a wrong sign for the kinetic terms cannot be valid for energies higher than
\begin{equation}
E\le 3 {\rm MeV}
\end{equation}
coming from the production of a gamma-ray background according to: ${\rm vacuum} \to 2\phi+ 2\gamma$ where the vacuum creates particles from nothing.
Of course, it seems difficult to believe that such a low energy effective theory with ghosts exists at all\cite{Cline:2003gs}.

\subsection{Dark energy difficulties}

Dark energy suffers from two major drawbacks. A very natural problem follows from the late appearance of cosmic acceleration. If this is due to a pure cosmological constant  then the instant when cosmic acceleration sets in only depends on known features of the radiation and matter eras. Indeed the decrease of the matter density with time is known and the instant when it becomes subdominant to the cosmological constant can be determined. On the other hand, once the energy density of dark energy becomes dynamical and evolves in time, one may wonder why acceleration starts so late in the history of the universe. This can be reformulated as a question: what  is so special about the present epoch that the matter density and the dark energy density are of the same order of magnitude. This is the coincidence problem. In the Ratra-Peebles class of potentials, such a coincidence requires to tune the value of the scale $M$. For exponential potentials, the initial condition of the field $\phi_{\rm ini}$ has to be carefully chosen.
Of course, in the Ratra-Peebles case one could argue that the scale $M$ should be derived from a more fundamental theory therefore leading to a solution of the coincidence problem. So far, no solution to the coincidence problem has ever been uncovered.

Another important difficulty springs from the role of quantum corrections\cite{Kolda:1998wq}. We have already seen that the value of the potential can be shifted by quantum corrections. As a result, dark energy is  a valid concept provided the cosmological constant problem has been solved, i.e. the minimal value of the potential  vanishes. In thawing models, this requires the existence of a minimum with a vanishing potential. In freezing models, the asymptotic value of the potential when the field rolls to very large values has to vanish too. We have seen that no intrinsically field theoretical explanation
has been given to this puzzle so far. It might be that the vanishing of the minimal value of the potential has an anthropic origin.
If this could be  ascertained, dark energy with specific potential could lead to an explanation of cosmic acceleration due to a tiny vacuum energy density.

Unfortunately the dark energy mechanism is also jeopardised by the smallness of the dark energy mass $m_\phi \approx H_0$. This is such a tiny scale compared to particle physics ones that it can be drastically shifted by quantum corrections due to the decoupling of heavy particles of mass $M$
\begin{equation}
\delta m_\phi^2\approx  \frac{\beta^2 M^4}{16\pi^2 m_{\rm Pl}^2}
\end{equation}
where  the coupling $\beta\equiv \alpha_\phi$ for scalar-tensor theories. The correction term is too large as soon as $M\ge 10^{-3}$ eV corresponding to neutrino masses if the coupling $\beta=O(1)$. Of course corrections to the dark energy mass coming from standard model particles are way too large unless
\begin{equation}
\beta \le 10^{-40}
\end{equation}
This would lead to an almost complete decoupling of dark energy from matter. In all other cases, stabilising the mass of the dark energy field
is a very hard problem.

\subsection{Pseudo-Goldstone bosons}

Goldstone bosons appear when a global symmetry is broken at a scale $f$. Typically, consider a $U(1)$ global symmetry
with a complex scalar field $\Phi$ acquiring a vev $<\Phi>= f$ at the minimum of its potential $V(\Phi)$. Writing
\begin{equation}
\Phi= f e^{i\phi/\sqrt 2 f}
\end{equation}
the boson $\phi$ is a Goldstone boson with normalised kinetic terms. The Goldstone field has no potential and is therefore massless.
Moreover the model has a residual shift symmetry $\phi\to \phi +c$ where $c$ is real corresponding to the original $U(1)$ global symmetry. This implies that couplings to fermions of the type
\begin{equation}
{\cal L}_{\rm int}= \beta \phi \bar \psi \psi
\end{equation}
are forbidden.
A potential for  the Goldstone boson can be obtained by a small breaking of the $U(1)$ symmetry
\begin{equation}
{\cal L}_{\rm breaking}= \mu^4 \frac{\Phi+\bar \Phi}{2f}
\end{equation}
where $\mu\ll f$. This leads to the potential
\begin{equation}
V(\phi)= \mu^4 \cos \frac{\phi}{\sqrt 2 f}
\end{equation}
This model is of the thawing type\cite{Burgess:2004ib}. Initially the field starts away from the minimum of the potential and stays there due to the Hubble friction.
Later, it converges to the minimum of the potential. Cosmic acceleration implies that
\begin{equation}
\mu \sim 10^{-3} {\rm GeV}
\end{equation}
The mass of the pseudo Goldstone boson is then
\begin{equation}
m_\phi^2\approx \frac{\mu^4}{2 f^2}
\end{equation}
Requiring that $m_\phi\approx H_0$ leads to
\begin{equation}
f\approx m_{\rm Pl}
\end{equation}
Corrections to the  mass of the pseudo Goldstone boson arise from derivative couplings of the field $\Phi$ to fermions. The lowest order term is
\begin{equation}
{\cal L}_{\rm int}=  \frac{\bar \Phi\partial^2 \Phi}{f^3} \bar \psi \psi
\end{equation}
This non-renormalisable operator is suppressed by the Planck scale as the model must be valid up to this scale to accommodate $f\approx m_{\rm Pl}$.
To leading order, this gives an interaction term for the pseudo Goldstone mode
\begin{equation}
{\cal L}_{\rm int}\approx  \frac{\partial^2 \phi}{f^2 }\bar \psi \psi
\end{equation}
The mass correction due to this interaction is of order
\begin{equation}
\delta m_\phi^2 \approx \frac{m_\phi^4}{f^4}\frac{M^2}{16\pi^2}
\end{equation}
Of course this is a tiny correction. Hence pseudo Goldstone boson solve the mass problem of dark energy.

An explicit realisation of the Goldstone idea can be obtained using neutrino physics\cite{Barbieri:2005gj}. Indeed consider a family of three right handed neutrinos
$N_i$ whose Majorana mass matrix is due to the interaction with 6 scalar fields $\Phi_{ij}$ according to
\begin{equation}
{\cal L}_N= \frac{1}{2} \lambda_{ij} \Phi_{ij} N_iN_j
\end{equation}
The model is invariant under a $U(1)^3$ symmetry obtained by changing the phases of the right handed neutrinos. When the couplings $\lambda_{ij}=0$ the model is invariant under $U(6)$ corresponding to changing the phases of the six scalars. Now when $\Phi_{ij}$ acquires a vev $<\Phi_{ij}>= f_{ij}$ the six Goldstone bosons separate into three true Goldstone bosons and three pseudo-Goldstone bosons due to the Majorana interaction terms. A potential for the pseudo-Goldstone is generated at one loop by the Coleman-Weinberg mechanism
\begin{equation}
V= \frac{1}{32\pi^2}{\rm tr} ( MM^\dagger MM^\dagger \ln (\frac{E^2} {M^\dagger M}))
\end{equation}
where $M_{ij}= \lambda_{ij} f_{ij} e^{i\phi_{ij}/\sqrt 2 f_{ij}}$ with no summation involved and $E$ is the limit of  validity of the model. The fields $\phi_{ij}$ are the six Goldstone bosons.
This  gives a potential to the three pseudo Goldstone bosons of the form
\begin{equation}
V(\phi)\approx \mu^4 \cos (\frac{\phi}{\sqrt 2 f})
\end{equation}
where $\mu^4 \sim M^4$. Now neutrino oscillation experiments are compatible with $\mu \sim 10^{-3}$ eV providing a possible link between neutrino physics and dark energy.

\section{Scalar-Tensor Theories}

Many well-motivated models lead to scalar-tensor theories where a scalar field couples to matter on par with
the gravitational field. Another type of models, the $f(R)$ theories, are also scalar-tensor theories.

\subsection{Jordan vs Einstein}

The action governing the dynamics of the field $\phi$ in a scalar-tensor theory is
of the general form
\begin{equation}
S=\int d^4x\sqrt{-g}\left\{\frac{m_{\rm Pl}^2}{2}{
R}-\frac{1}{2}(\partial\phi)^2- V(\phi)\right\} - \int d^4x{\cal
L}_m(\psi_m^{(i)},\tilde g_{\mu\nu})\,, \label{action}
\end{equation}
where
$g$ is the determinant of the metric $g_{\mu\nu}$, ${ R}$ is
the Ricci scalar and $\psi_m^{(i)}$ are various matter fields
labeled by $i$. A key ingredient of the model is the conformal
coupling of $\phi$ with matter particles. More precisely, the
excitations of each matter field $\psi_m^{(i)}$ follow the
geodesics of a metric $\tilde g_{\mu\nu}$ which is related to the
Einstein-frame metric $g_{\mu\nu}$ by the conformal rescaling
\begin{equation}
\tilde g_{\mu\nu}=A^2(\phi)g_{\mu\nu}\,,
\label{conformal}
\end{equation}
The Klein Gordon equation is modified due to the coupling of the scalar field to matter
\begin{equation}
\Box \phi= -\alpha_\phi T + \frac{dV}{d\phi}
\end{equation}
where $T$ is the trace of the energy momentum tensor $T^{\mu\nu}$ and
the coupling of the scalar field to matter is defined by
\begin{equation}
\alpha_\phi= \frac{d\ln A}{d\kappa_4 \phi}
\end{equation}
This is equivalent to the usual Klein-Gordon equation with  the effective potential
\begin{equation}
V_{eff}(\phi) = V(\phi) - A(\phi)T
\label{veff}
\end{equation}
Matter is not conserved anymore. Indeed the fact that the scalar field couples to matter implies that
\begin{equation}
D_\mu T^{\mu\nu}= \alpha_\phi \partial^\nu \phi  T.
\end{equation}
 This can be also seen by looking at the Lagrangian involving matter and the scalar field.
Let us first examine the effect of the scalar field on couplings and masses. The scalar field does not couple to gauge field
\begin{equation}
S_{\rm gauge}= -\frac{1}{4g^2} \int {\rm d}^4 x \sqrt{-g} F_{\mu\nu} F^{\mu\nu}.
\end{equation}
On the other hand, the scalar field couples to both scalars and fermions with a mass
\begin{equation}
m(\phi)=A(\phi) m_0
\end{equation}
where $m_0$ is the bare mass in the Lagrangian. This leads to the possibility of varying particle masses.

In the cosmological context
\begin{equation}
T= -\rho+ 3p
\end{equation}
implying that the scalar field does not couple to radiation. Moreover the conservation equation becomes
\begin{equation}
\dot\rho+ 3H( \rho+p) = \alpha_\phi \dot \phi (\rho-3p)
\end{equation}
In the case of cold dark matter, this is the equation we have used in the coupled dark energy case where $\alpha_\phi$ is constant.
The effective potential depends on the amount of matter
\begin{equation}
V_{eff}(\phi) = V(\phi) +A(\phi)\rho_m
\end{equation}
This has spectacular consequences for chameleon fields.

Matter follows geodesics which deviate from the ones in ordinary gravity
\begin{equation}
\frac{d^2x^i}{d\tau^2}
+\Gamma^{i}_{jk}\frac{dx^j}{d\tau}\frac{dx^k}{d\tau} +\alpha_\phi\kappa_4
\frac{\partial \phi}{\partial x_i}=0
\end{equation}
in terms of the proper time $\tau$ and the Christoffel symbols
calculated with the metric $g_{\mu\nu}$.
In the Newtonian case
where
\begin{equation}
ds^2=-(1+2\Phi_N) dt^2+ (1-2\Phi_N)dx^idx_i
\end{equation}
and assuming that $\alpha_\phi$ is slowly varying along the particle trajectory,
this  reduces to
\begin{equation}
\frac{d^2 x^i}{dt^2}=- \partial^i (\Phi_N +\alpha_\phi\kappa_4
\phi)
\end{equation}
This can be interpreted as the motion of a particle in
 the effective gravitational potential which is
\begin{equation}
\tilde \Psi= \Phi_N + \alpha_\phi \kappa_4\phi
\end{equation}
This is the potential acting on a single particle. The scalar field induces a modification of gravity.

Modifications of gravity are drastically illustrated for models with $V=0$, i.e. a massless scalar field and $\rho=m_0\delta^{(3)}$ for a massive point particle.
In this case the static Klein-Gordon equation reads
\begin{equation}
\Delta \phi= \alpha_\phi \kappa_4 m_0\delta^{(3)}
\end{equation}
whose solution is
\begin{equation}
\phi= -\frac{\alpha_\phi \kappa_4}{4\pi r}
\end{equation}
implying that
\begin{equation}
\tilde\Psi= - (1+2\alpha_\phi^2)\frac{G_N}{r}
\end{equation}
leading to a modification of Newton's law depending on the coupling $\alpha_\phi$
\begin{equation}
G_{N,eff}=\gamma G_N,\
\end{equation}
where $\gamma$ is one of the Eddington post-Newtonian parameters
\begin{equation}
\gamma=1+ 2\alpha_\phi^2.
\end{equation}
These results have been obtained in the Einstein frame where the Einstein-Hilbert term in the action is
not modified\cite{Damour:1992we}.

There is another representation of the same theory in the Jordan frame obtained by a Weyl rescaling amounting to considering that $\tilde g_{\mu\nu}$ is the dynamical field representing gravity. Using the transformation law
\begin{equation}
R= A^{2}(\phi)(\tilde R+6\tilde g^{\mu\nu} \tilde D_\mu \tilde D_\nu \ln A(\phi)- 6\tilde g^{\mu\nu} \tilde D_\mu\ln A(\phi)\tilde D_\nu\ln A(\phi))
\end{equation}
the Jordan frame action reads

\begin{equation}
S=\int d^4x\sqrt{-\tilde g}\left\{\frac{M_{Pl}^2}{2A^2(\phi)}\tilde {
R}-\frac{1}{2A^2(\phi)}(1-6\alpha_\phi^2)(\partial\phi)^2- \frac{V(\phi)}{A^4(\phi)}\right\} - \int d^4x{\cal
L}_m(\psi_m^{(i)},\tilde g_{\mu\nu})\,, \label{action}
\end{equation}
Notice that in this frame matter is conserved. The Jordan frame is such that matter couples minimally to gravity. The modification of gravity
appears due to the $A^2(\phi)$ in the Einstein-Hilbert term. In particular, the measured Newton constant in Cavendish experiments is
\begin{equation}
\tilde G_{N}=(1+2\alpha_\phi ^2) A^2(\phi) G_N
\end{equation}
The coupling constant $\alpha_\phi$ is strongly constrained by the Cassini experiment\cite{Bertotti:2003rm}
\begin{equation}
\alpha_\phi^2 \le 10^{-5}
\end{equation}
where in most cases $A(\phi_0)\approx 1$.
If the scalar field does not couple to baryons as in coupled dark energy, the constraint is relaxed.

\subsection{Violation of the equivalence principle}

Of course, gravitational experiments are not
carried out on microscopic particles but on macroscopic objects
composed of many atoms. The mass of a particular atom can be
decomposed as\cite{Damour:1992we}
\begin{equation}
m_{_{\rm ATOM}}\simeq M \Lambda _{_{\rm QCD}}+ \sigma' \left(N+Z\right)
+ \delta' \left(N-Z\right) + a_3
\alpha_{_{\rm QED}} E_{\rm A} \Lambda_{_{\rm QCD}} \, ,
\label{atom}
\end{equation}
where $\Lambda _{_{\rm QCD}}\simeq 180\ \mbox{MeV}$ is the QCD scale,
$N$ the number of neutrons and $Z$ the number of protons. The quantity
$M$ can be written $M= (N+Z) + E_{_{\rm QCD}}/\Lambda_{\rm QCD}$ where $
E_{_{\rm QCD}}$ is the strong interaction contribution to the nucleus
binding energy. The number $E_{\rm A}$ is given by $E_{\rm A}=
Z(Z-1)/(N+Z)^{1/3}$ and the quantity $a_3\alpha _{_{\rm QED}}\Lambda
_{_{\rm QCD}}E_A$ represents the Coulomb interaction of the nucleus
where $a_3\alpha_{_{\rm QED}}\simeq 0.77\times 10^{-3}$. Finally the
coefficients $\delta'$ and $\sigma'$ depend on the constituent masses
and can be expressed as
\begin{eqnarray}
\label{eq:mess1}
\sigma'&=& \frac{1}{2}(m_{\rm u}+m_{\rm d})(b_{\rm u}+b_{\rm d})
+\frac{\alpha_{_{\rm
QED}}}{2}(C_{\rm n}+C_{\rm p}) +\frac{1}{2} m_{\rm e}\, , \\
\label{eq:mess2}
\delta'&=&\frac{1}{2}(m_{\rm u}-m_{\rm d})(b_{\rm u}-b_{\rm d})
+\frac{\alpha_{_{\rm QED}}}{2} (C_{\rm n}-C_{\rm p}) -\frac{1}{2} m_{\rm
e}\, ,
\end{eqnarray}
where $m_{\rm u}\sim 5~{\rm MeV}$, $m_{\rm d} \sim {10}~{\rm MeV}$. The
constants appearing in Eqs.~(\ref{eq:mess1}) and~(\ref{eq:mess2}) are
given by: $b_{\rm u}+b_{\rm d}\simeq 6$, $b_{\rm u}-b_{\rm d}\sim 0.5$,
$C_{\rm p} \alpha_{_{\rm QED}} \simeq 0.63~{\rm MeV}$, $C_{\rm
n}\alpha_{_{\rm QED}}\sim -0.13~{\rm MeV}$. This implies that
$\sigma'/\Lambda_{_{\rm QCD}}\simeq 3.8 \times 10^{-2}$ and
$\delta'/\Lambda_{_{\rm QCD}} \simeq 4.2\times 10^{-4}$.

\par

The fact that $m_{\rm u}, m_{\rm d}$ and $m_{\rm e}$ are dark energy-dependent
quantities imply that the coefficients $\sigma '$ and $\delta '$, and
hence $m_{_{\rm ATOM}}$, are now $\phi$-dependent quantities.
We are now in a position to estimate the typical
gravitational coupling of an atom. From the previous considerations,
one obtains
\begin{equation}
\kappa_4 \alpha_{_{\rm ATOM}} = \frac{N+Z}{M} \frac{\partial}{\partial
\phi}\left(\frac{\sigma'}{\Lambda_{_{\rm QCD}}}\right) +\frac{N-Z}{M}
\frac{\partial}{\partial \phi}
\left(\frac{\delta'}{\Lambda_{_{\rm QCD}}}\right)
\end{equation}
where the variations of $\sigma'$ and $\delta'$ read
\begin{eqnarray}
\frac{\partial}{\partial\phi}\left(\frac{\sigma'}
{\Lambda_{_{\rm QCD}}}\right)&=&
\frac{\kappa_4}{2\Lambda_{_{\rm QCD}}} \left [(b_{\rm u}+b_{\rm d})
\left( m_{\rm u} +m_{\rm d}\right)
+ m_{\rm e}\right ] \alpha_\phi
\nonumber \\
\frac{\partial}{\partial\phi}
\left(\frac{\delta'}{\Lambda _{_{\rm QCD}}}\right)&=&
-\frac{\kappa_4}{2\Lambda _{_{\rm QCD}}} \left[\left(b_{\rm u}-b_{\rm d}\right)
\left( m_{\rm u} -m_{\rm d}\right)+
 m_{\rm e}\right ]\alpha_\phi
\nonumber \\
\end{eqnarray}
They are proportional to the microscopic coupling  $\alpha_\phi$. On the other hand, each coupling to each type of atom is different due to the
different numbers of protons and neutrons. As a result, we find that each atom couples as
\begin{equation}
\alpha_A= \beta_A \alpha_\phi
\end{equation}
where $\beta_A$ can be deduced from the previous formulae
\begin{equation}
\beta_A= \frac{N+Z}{M}\frac{(b_u+b_d)(m_u+m_d)+m_e}{2\Lambda_{\rm QCD}}+ \frac{Z-N}{M}\frac{(b_u-b_d)(m_u-m_d)+m_e}{2\Lambda_{\rm QCD}}
\end{equation}
Here the violation of the equivalence principle is measured by
\begin{equation}
\eta_{AB}= 2\frac{\vert a_A-a_B\vert}{a_A+a_B}
\end{equation}
where $a_A$ is the acceleration of the atom $A$ in the gravitational field of a body $E$, it is given by
\begin{equation}
\eta_{AB}=\vert \beta_A- \beta_B\vert \beta_E \alpha_\phi^2
\end{equation}
Now the present limit on the violation of the equivalence principle is $\eta \le 10^{-12}$. For a typical atom $\beta=0.1$ this leads to  a  strong bound
\begin{equation}
\alpha_\phi  \le 10^{-5}
\end{equation}
which is much stronger than the Cassini bound.
This bound applies to dark energy models as the mass of the dark energy field $m_\phi \approx H_0$ is almost zero (more precisely the range of the fifth force mediated by the scalar field is of the order of the Hubble horizon, i.e. much larger than the distances which are gravitationally probed).
As a result, one must explain why the coupling of dark energy is so weak. For thawing models of the pseudo-Goldstone type, this is natural as
the direct coupling to matter is forbidden by a  shift symmetry. For freezing models with  runaway potentials, this appears to be another
fine-tuning. Fortunately, a class of models known as chameleons do not suffer from these problems.

\subsection{Examples}

Let us give  non-trivial examples of scalar-tensor models. Consider the low energy theory of the Randall-Sundrum model with two branes and matter on the positive tension brane. At low energy,  the distance between the branes becomes a dynamical field $\phi$ according to
\begin{equation}
d=-l\ln (\tanh \frac{\kappa_4 \phi}{\sqrt 6})
\end{equation}
where $l^{-2}$ is the bulk Anti de Sitter curvature.
The coupling to matter is
\begin{equation}
A(\phi)= \cosh \frac{\kappa_4 \phi}{\sqrt 6}
\end{equation}
implying that\cite{Brax:2004ym}
\begin{equation}
\alpha_\phi= \frac{\tanh \frac{\kappa_4 \phi}{\sqrt 6}}{\sqrt 6}
\end{equation}
When the branes are far apart the coupling is small. It is equal to $ \alpha_\phi=1/\sqrt 6$ for close branes.

Another interesting class of models arises in supergravity when dark energy belongs  to a hidden sector
\begin{equation}
K= K_{\rm matter}+ K_{\rm DE}(\phi,\bar\phi), \ \ \ \ W= W_{\rm matter}+ W_{\rm DE}(\phi)
\end{equation}
The only interactions between the dark energy and matter sectors are gravitational.
In particular the mass of matter fermions becomes
\begin{equation}
m_{\rm fermion}(\phi)= e^{\kappa_4^2 K_{\rm DE}(\phi,\bar\phi)/2} m_{\rm fermion}
\end{equation}
The prefactor is nothing but the $A(\phi)$ factor in these models.
Noticing that the dark energy field is not canonically normalised ${\cal L}_{\rm kin}= K_{\phi\bar\phi} \partial_\mu\phi\partial^\mu \bar \phi$, this leads to\cite{Brax:2009kd}
\begin{equation}
\alpha_\phi= \kappa_4 \frac{K_\phi}{2\sqrt K_{\phi\bar\phi}}
\end{equation}
For canonically normalised fields $K_{\rm DE}= \phi\bar\phi$, this gives $\alpha_\phi= \frac{\kappa_4\phi}{2}$ which is of order one for runaway models where $\phi_0\approx m_{\rm Pl}$. For moduli fields with $K_{\rm DE}=-3m_{\rm Pl}^2 \ln (\kappa_4(\phi+\bar \phi))$ this is $\alpha_\phi=\frac{\sqrt 3}{2}$. In both cases, the models are ruled out unless the chameleon mechanism is at play.

In fact, a natural way of guaranteeing the vanishing of $\alpha_\phi$ along  $\phi=\bar\phi$  taken as the dark energy direction is to impose that\cite{Brax:2009kd}
\begin{equation}
K_{\rm DE}(\phi\,\bar\phi)= K_{\rm DE}(\phi-\bar\phi)
\end{equation}
The model has a shift symmetry $\phi\to \phi+c$ characteristic of pseudo-Goldstone bosons or axions. If dark energy can be derived from a
fundamental theory such as string theory, it is likely that the dark energy particle will be an axion\cite{Choi:1999xn}.

\subsection{Unimodular gravity}

Unimodular gravity\cite{Shaposhnikov:2008xb} is an interesting modification of general relativity where  the cosmological constant is not dynamical anymore, therefore solving the cosmological constant problem. The gravity Lagrangian becomes
\begin{equation}
S= \int {\rm d}^4x (\frac{R}{2\kappa_4^2} + M_0^4)
\label{Uni}
\end{equation}
where $\det ( - g)=1$. The cosmological constant $M_0^4$ is non-dynamical as it does not appear in the equations of motion. In Einsteinian gravity, the cosmological constant receives quantum field corrections due to the non-trivial nature of vacuum fluctuations. In unimodular gravity,  the  radiative corrections do not lead to the acceleration of the universe despite the possible existence of a large cosmological constant. The cosmological constant can be arbitrary and
has  no consequence on the dynamics of the universe. Of course, this possibility does not explain the observed acceleration of the late time universe. All it provides is a mechanism for the irrelevance of the quantum fluctuations of the vacuum.

In unimodular gravity, scale invariance is explicitly broken as $\det (-g)=1$. The breaking of scale invariance results in the effective appearance of a scale $M^4$ in the theory. This scale can be understood as the Lagrange multiplier enforcing the constraint
\begin{equation}
S= \int {\rm d}^4x [\sqrt{-g}\frac{R}{2\kappa_4^2} + M_0^4- M^4 (\sqrt{-g}-1))]
\label{Ein}
\end{equation}
Using the action (\ref{Ein}), the equations of motion give  $\det( -g )=1$ and
\begin{equation}
R_{\mu\nu}-\frac{g_{\mu\nu}}{2} R= -\kappa_4^2 M^4 g_{\mu\nu}
\end{equation}
Taking the covariant derivative of the latter equation, the Bianchi identity implies that $M$ is a constant. This scale  is not dynamical and is simply set by
the initial conditions of the theory. In the Einstein equations it acts as a
pure cosmological constant. Of course this $M^4$ term is independent of the vacuum energy $M_0^4$. In fact it is a boundary condition which has to be set in the early universe.

 Dark energy can be incorporated in unimodular gravity by defining the following scalar-tensor theory
\begin{equation}
S_{\rm uni}= \int {\rm d}^4x (f(\phi^i) {R} -h_{ij}(\phi^k) \partial_\mu\phi^i\partial_\nu\phi^j -  V(\phi^i))
\end{equation}
where the mass dimensions are such that
are $[\phi^i]=1,\ [f]=2,\ [h_{ij}]= 0,\ [V]=4$, and  the mass dimensions coincide with the scale dimensions under the rescaling $x^\mu\to \lambda^{-1} x^\mu$.
Unimodularity  can be enforced using a Lagrange multiplier
\begin{equation}
S= \int {\rm d}^4x (\sqrt{-g}[(f(\phi^i) {R} -h_{ij}(\phi^k) \partial_\mu\phi^i\partial_\nu\phi^j-  V(\phi^i))] -M^4 (\sqrt {-g}-1))
\end{equation}
As in the case of pure unimodular gravity, one can show that $M$ is a constant.
The equations of motions enforce $\det ({-g}) =1$ and also two sets of equations which coincide with
the  Einstein  and the Klein-Gordon equations of  the fully covariant Lagrangian
\begin{equation}
{\cal L}= \sqrt{-g}[(f(\phi^i) {R} -h_{ij}(\phi^k) \partial_\mu\phi^i\partial_\nu\phi^j - V(\phi^i) -M^4]
\end{equation}
Hence the dynamics of the theory are determined by this scalar-tensor theory.

Homogeneity allows us to single out one field $\Phi \equiv \phi^1$ and define $z^i=\phi^i/\phi^1$. Therefore the coupling functions become
\begin{equation}
f(\phi^i)= \Phi^2 \tilde f(z^i),\ h_{ij}(\phi^k)= h_{ij}(z^k), \ V(\phi^i)= \Phi^4 \tilde V(z^i)
\end{equation}
Vacua of the theory are defined by minima of the potential $\partial_{\phi^i}V=0$ implying that $\tilde V=0$ and $ \partial_{z^i} \tilde V=0$.
Notice that the vev of $\Phi$ is not determined at all: it is a Goldstone direction preserving the vacuum of the theory. In fact $\Phi$ is only a pseudo -Goldstone mode as the breaking of scale invariance in unimodular  gravity lifts the vacuum energy density and gives a potential to $\rho$. The pseudo-Goldstone mode is a natural candidate for dark energy.

Let us write the effective action along the Goldstone direction. Defining a constant $\gamma$ such that
$
h_{ij}( <z^i>) \partial_\mu \phi^i\partial^\mu \phi^j= 6 \gamma \partial_\mu \Phi\partial^\mu \Phi
$
we find that the effective action for the Goldstone mode is
\begin{equation}
{\cal L_{\rm eff}}= \sqrt{-g}[(f_0\Phi^2  {R} -  6\gamma  \partial_\mu \Phi\partial^\mu \Phi -M^4]
\end{equation}
where $ \tilde f(<z^a>)=f_0$.
So far, we have not included matter fields. They can easily be introduced by  adding their action to the previous model
where the  matter fields $\psi_m$ couple to the metric $g_{\mu\nu}$.

In  order to analyse the gravitational and cosmological properties
of the model it is easier to go to the Einstein frame with
\begin{equation}
A^{2}(\Phi)= (2f_0  \kappa_4^2 \Phi^2)^{-1}
\end{equation}
In the Einstein frame the action reads
\begin{equation}
 {\cal S_{\rm eff}}= \int d^4 x \sqrt{-g}[\frac  { R}{2\kappa_4^2} -  6(\gamma+f_0) A^{2}(\Phi) \partial_\mu \Phi\partial^\mu \Phi -M^4 A^{4}(\Phi)]
\end{equation}
It is useful to normalise
\begin{equation}
\Phi=\frac{1}{\sqrt {2f_0} \kappa_4} e^{\kappa_4\sqrt f_0 \phi/\sqrt{6(\gamma+f_0)}}
\end{equation}
leading to the effective action
\begin{equation}
 S= \int {\rm d}^4 x \sqrt{-g}[\frac  {R}{2\kappa_4^2} - \frac{1}{2}  \partial_\mu \phi\partial^\mu \phi -M^4 e^{-4\kappa_4\frac{\sqrt{f_0}}{\sqrt{6(\gamma+f_0)}}\phi}] + S_m (\psi_m, e^{-2\kappa_4\frac{\sqrt{f_0}}{\sqrt{6(\gamma+f_0)}}\phi}\tilde g_{\mu\nu})
\end{equation}
We recognise a coupled dark energy model and we can identify
\begin{equation}
\lambda=4\alpha_\phi
\end{equation}
where
\begin{equation}
\alpha_\phi= \frac{\sqrt{f_0}}{\sqrt{6(\gamma+f_0)}}
\end{equation}
 The equation of state in this model is given by
\begin{equation}
1+w_\phi= \frac{16\alpha_\phi^2}{3}
\end{equation}
The Cassini bound implies that $\alpha_\phi^2$ must be less than $10^{-5}$. Therefore, dark energy in this model  is not very different  from
a pure cosmological constant.

\section{Chameleons and f(R) Gravity}

\subsection{Chameleons}

We have seen that the very small mass scale of dark energy wreaks havoc when coupled to matter as it leads to a strong violation of the equivalence principle if $\alpha_\phi=O(1)$. This conclusion is modified for certain scalar-tensor theories called chameleon theories\cite{Khoury:2003rn,Brax:2004qh}. Recall that in the presence of matter the effective potential of scalar-tensor theories is
\begin{equation}
V_{eff}(\phi) = V(\phi) +A(\phi)\rho_m
\end{equation}
where $\rho_m$ is the ambient matter density. In the following we shall assume that $V(\phi)$ is a decreasing function as befitting freezing models where the dark energy field rolls towards large values. An interesting phenomenon occurs when $A(\phi)$ is an increasing function: the effective potential acquires an energy density minimum satisfying
\begin{equation}
\frac{dV}{d\phi}\vert_{\phi_{\rm min}}= -\alpha_\phi A(\phi_{\rm min}) \frac{\rho_m}{m_{\rm Pl}}
\end{equation}
As $\rho_m$ increases, the position of the minimum moves to the origin. At the minimum, the dark energy field has a mass
\begin{equation}
m^2_\phi= \frac{d^2 V}{d\phi^2} + \frac{d^2 A}{d\phi^2} \rho_m
\end{equation}
This mass depends on the density $\rho_m$. In a very dense environment, the mass can be large enough to shorten the range of the force
mediated by the scalar field below the observable threshold. Gravity experiments have tested Newton's law down to a scale of about $0.1$ mm. The range becomes shorter and therefore unobservable provided $m_\phi\ge 10^{-3}$ eV\cite{Adelberger:2004ct}.
In a sparser environment such as the solar system, the mass is generically too small to evade strong deviations of planet trajectories.
Another effect must be at play called the thin shell effect.

Consider a large spherical object of radius $R$ and density $\rho_c$. Outside the body, the density is much smaller $\rho_\infty$. In this environment, the profile of the scalar field is non-trivial. Inside the body, the field is constant at a value $\phi_c$ corresponding to the minimum
of the effective potential with a density $\rho_c$. The field has a rapid variation over a shell of size $\Delta R$ below the surface of the body.  Far away from the body the field reaches a constant value $\phi_\infty$ corresponding to the density $\rho_\infty$. The solution outside the body is independent of the details of the potential and reads
\begin{equation}
\phi (r)= -\left (\frac{\alpha_\phi(\phi_\infty)}{4\pi m_{\rm Pl}}\right ) \left (\frac{R^3-R_s^3}{R^3}\right ) \frac{M e^{-m_\infty (r-R)}}{r} +\phi_\infty
\end{equation}
where $M$ is the mass of the spherical body and $R_s=R-\Delta R$.
Over distances shorter than the large range of the scalar field force outside the body, this corresponds to a modification of Newton's potential by a factor
\begin{equation}
\frac{\delta \tilde \Psi}{\tilde \Psi}(r)= 2 \alpha_\phi^2 \left ( \frac{R^3-R_s^3}{R^3}\right ).
\end{equation}
For a point-like object where $R\to 0$ and no thin-shell, we retrieve the $2\alpha_\phi^2$ correction. For larger objects, the correction is small when the size of the shell is small
\begin{equation}
\frac{\Delta R}{R}= \frac{\phi_\infty -\phi_c}{6\alpha_\phi(\phi_\infty) m_{Pl} \Phi_N (R)}.
\end{equation}
This is the case when the thin shell condition is satisfied
\begin{equation}
(\phi_\infty -\phi_c)\ll 6\alpha_\phi (\phi_\infty) m_{Pl} \Phi_N (R).
\end{equation}
When a thin shell exists for large bodies, the field is called a chameleon.

Thin shells do not exist for all couplings and potentials. If both are exponential,  a thin shell cannot be obtained for planets in the solar system.
A class of models with nice properties is obtained with
\begin{equation}
V(\phi)= M_0^4 +\frac{M^{4+n}}{\phi^n}
\end{equation}
They correspond to Ratra-Peebles potentials corrected by a pure cosmological constant.
The coupling is chosen to be
\begin{equation}
A(\phi)= e^{\alpha_\phi \kappa_4 \phi}
\end{equation}
Cosmic acceleration requires that $M_0\sim 10^{-3}$ eV. The thin shell condition for test objects used in laboratory experiments implies that
$M\sim M_0$. This is a numerical coincidence which is highly intriguing. We will come back to the type of potentials leading to chameleons when we discuss $f(R)$ models.

Cosmologically,  chameleons have a very simple behaviour. To simplify the discussion, we assume that $\alpha_\phi=O(1)$ and almost constant. Then we can derive a lower bound on the chameleon mass
\begin{equation}
m^2\ge 3 \alpha_\phi^2 H^2
\end{equation}
where we assume that $\phi\ll m_{\rm Pl}$ in cosmology. This is always the case in known models.
The fact that the mass at the minimum is always greater than the Hubble rate implies that the minimum is a cosmological attractor. After a sufficiently long time, the chameleon converges to the minimum. Notice that the minimum evolves as the matter density decreases.
Prior to converging to the minimum, the chameleon field can either undershoot or overshoot.
Undershooting corresponds to an initial chameleon $\phi_i\gg \phi_{\rm min}$. During this period the potential is negligible and the chameleon obeys
\begin{equation}
\ddot \phi + 3H \dot \phi= \frac{\alpha_\phi}{m_{\rm Pl}} T^\mu_\mu
\end{equation}
During the radiation era, the trace of the energy momentum tensor is almost zero implying that the field is stuck at its initial value. It would only start moving during the matter era. This is only part of the story as one can show that every time a particle becomes non-relativistic the trace of the energy momentum tensor becomes non-zero during a Hubble time. As it happens, the field receives a kick and moves rapidly before grinding to a halt again. The cumulative effect of these multiple kicks is a variation
\begin{equation}
\Delta \phi\approx \alpha_\phi m_{\rm Pl}
\end{equation}
Overshooting is the opposite situation where $\phi_i\ll \phi_c$. In this case, the chameleon starts very high along the potential and starts tumbling down very fast. During this free fall evolution, the field does not feel the potential and eventually stops due to the Hubble friction
\begin{equation}
\phi_{\rm stop}= \phi_i + \sqrt {6 \Omega_{\phi }^i} m_{\rm Pl}
\end{equation}
where $\Omega^i_\phi$ is the initial density fraction of the chameleon. After this the chameleon is in an undershoot situation.

A variation of the chameleon of order $\Delta \phi$ implies a variation in the particle masses of order
\begin{equation}
\frac{\Delta m_f}{m_f}= \alpha_\phi \frac{\Delta \phi}{m_{\rm Pl}}
\end{equation}
Constraints from Big Bang Nucleosynthesis imply that this variation should not be larger than 10\%
between BBN and now.
If the chameleon is overshooting during BBN, its variation is bounded by $\sqrt {6 \Omega_{\phi }^i} m_{\rm Pl}$ leading to a bound
on the initial fraction $\Omega_\phi^i\le 10^{-2}$. Generically,  the field does not overshoot during BBN. It stops before  BBN and then undershoots.
If the chameleon is undershooting during BBN, then it receives a dangerous kick due to the electron becoming non-relativistic and leading to a variation of order $0.3 m_{\rm Pl}$. This is excluded. Hence the chameleon must have reached the vicinity of the minimum by BBN in which case the
electron kick is non-operative. This requires an initial value of order $\phi_i\le  m_{\rm Pl}$. This large value of the field can be obtained if the chameleon is initially overshooting with $\Omega^i_\phi\le 1/6$.
If the field is around the minimum during BBN then the subsequent evolution of the chameleon is negligible compared to the Planck scale leading to no problems with BBN. In conclusion, chameleon fields are generically required to be close to the minimum during BBN.

\subsection{f(R) models}

A class of modified gravity models turns out to be equivalent to certain scalar-tensor theories\cite{Sotiriou:2008rp}. Let us consider the theory defined by
\begin{equation}
S= \int d^{\rm 4 } x \sqrt{-g} \frac{f(R)}{2\kappa_4^2}
\end{equation}
where $f$ is an arbitrary function. These models extend the usual Einstein-Hilbert action involving high order terms with more than 2 derivatives.
It is more convenient to introduce a field  $\lambda$ and write
\begin{equation}
S=\frac{1}{2\kappa_4^2}\int {\rm d}^4 x \sqrt{-g}\left [ f(\lambda) + f'(\lambda) (R-\lambda)\right ]
\end{equation}
As long as $f''(\lambda) \ne 0$, this leads to $R=\lambda$ and the appropriate $f(R)$ action. The field $\lambda$ has no kinetic terms. The action is a particular form of a scalar-tensor theory with  $A^2= 1/f'(\lambda)$. In the Einstein frame, the model becomes
\begin{equation}
S= \int {\rm d^4} x \sqrt{-g}  \left [\frac{R}{2\kappa_4^2}- \frac{1}{2} \partial_\mu\phi \partial^\mu\phi -V(\phi)\right ]
\end{equation}
where
\begin{equation}
f'(\lambda)= e^{-\sqrt{\frac{2}{3}}\kappa_4 \phi}
\end{equation}
and
\begin{equation}
V(\phi)= \frac{\lambda f'(\lambda)-f(\lambda)}{2\kappa_4^2 (f'(\lambda))^2}
\end{equation}
The coupling is fixed
\begin{equation}
\alpha_\phi=\frac{1}{\sqrt 6}.
\end{equation}
As a result, the only way $f(R)$ theories can pass the gravitational tests is to impose that they must be chameleon theories in disguise.
This gives strong constraints on the shape of the potential as it must lead to the thin shell behaviour. Equivalently this restricts the possible functions $f(R)$.

Let us analyse the possible deviations of $f(R)$ models compared to a pure cosmological constant.
First of all, there is a very mild constraint coming from the requirement that there should exists a matter era which is then followed by an accelerated epoch. This is realised provided\cite{Amendola:2007nt}
\begin{equation}
m=\frac{Rf_{RR}}{f_R}
\end{equation}
is less than 0.1.
Another set of much more stringent constraints follows from the thin shell condition\cite{Brax:2008hh}.
Defining
\begin{equation}
\Phi= f'(\lambda)
\end{equation}
The equations of motions derived from Einstein equations are
\begin{equation}
-\frac{\Phi_{\eta\eta}}{a^2} -2 \frac{a_\eta^2}{a^4}\Phi_\eta= \frac{\kappa_4^2}{3}\left ( T + 2\Phi^3 V\right)
\end{equation}
and
\begin{equation}
3\frac{a_\eta^2}{a^4}=\frac{\kappa_4^2 \rho_m}{\Phi} + \kappa_4^2 \Phi V -3 \frac{a_\eta}{a^3} \frac{\Phi_\eta}{\Phi}
\end{equation}
interpreted as the Klein-Gordon and the Friedmann equations in conformal time
\begin{equation}
ds^2=a^2(\eta)(-d\eta^2 + dx^2)
\end{equation}
We will derive a constraint on the apparent equation of state  from these equations.

When observations are carried out, it is generally assumed that particle physics processes which happened in far away objects are identical to the ones we observe on earth. This selects the Jordan frame as the physical frame to interpret data. The equation of state is then inferred by identification in the Friedmann equation assuming that the dark energy fluid is conserved. This amounts to writing
\begin{equation}
H^2\equiv \frac{a_\eta^2}{a^4}= \kappa_4^2{3\Phi_0} (\rho_m + \rho_{\rm DE})
\end{equation}
where
\begin{equation}
\kappa_{4\rm eff}^2= \frac{\kappa_4^2}{\Phi_0}
\end{equation}
is the apparent gravitational coupling.
This apparent  Friedmann equation defines $\rho_{DE}$.
Conservation of matter defines the dark energy equation of state
\begin{equation}
\rho_{{\rm DE}, \eta}= -3 \frac{a_\eta}{a} (1+ w_{\rm DE}) \rho_{\rm DE}
\end{equation}
With these identifications the equation of state can be obtained
\begin{equation}
(1+w_{\rm DE})\Omega_{\rm DE}=\frac{\Phi_{\eta\eta}}{3\Phi a^2H^2}-2 \frac{\Phi_{\eta}}{\Phi aH}+ (\frac{\Phi_0}{\Phi}-1)\Omega_m
\end{equation}
The right hand side is of order of the variation of $\phi$  in Hubble time units. Denoting by $\Delta \phi$ the time variation of $\phi$ during one Hubble time, we find
\begin{equation}
\vert (1+w_{\rm DE})\Omega_{\rm DE}\vert = O(\frac{\alpha_\phi \Delta \phi}{m_{\rm Pl}})
\end{equation}
Imposing that objects such as the sun have a thin shell leads to an upper bound on the spatial variation of the scalar field
\begin{equation}
\frac{\alpha_\phi(\phi_\infty- \phi_c)}{m_{\rm Pl}} \le \frac{U}{3}
\end{equation}
where $U$ is Newton's potential of the inhomogeneity to which the object belongs (typically a galaxy cluster).
Using the fact that the field at infinity increases with time $\phi_c<\phi (t)< \phi (t_0)$, we find that $\Delta \phi= \phi_\infty (t_0)-\phi_\infty (t) \le \phi_\infty (t_0)-\phi_c$ and therefore
\begin{equation}
\vert (1+w_{\rm DE})\Omega_{\rm DE}\vert\le \frac{U}{3}
\end{equation}
The largest structures in the universe have $U=O(10^{-4})$ leading to an upper bound
\begin{equation}
\vert (1+w_{\rm DE})\Omega_{\rm DE}\vert \le 10^{-4}
\end{equation}
This is a loose bound which can be supplemented by explicit calculations using constraints from laboratory gravitational tests such as the Eotwash experiment. In all known cases, the equation of state is bounded and lies extremely close to $-1$. The fact that $f(R)$ theories cannot be distinguished from a pure cosmological constant at the background level is nicely counterbalanced by the fact that perturbation theory and the growth of structures are  very different in chameleon theories (hence for viable $f(R)$ models)

\subsection{Structure formation}

It is actually easy  to analyse the growth of structures in $f(R)$ models using the equivalent scalar-tensor theory. In this section we will present results which are valid for all chameleon models.
We assume that the background is such that the chameleon sits at the minimum of its effective potential. As we have seen, this must be the case since BBN and therefore is the case of interest when structures form in the matter era. We also assume that the chameleon energy density is subdominant, i.e. we take $z\ge O(1)$. In the linear regime,
we are interested in the density contrast
\begin{equation}
\tilde \delta = \frac{\delta \tilde \rho} {\tilde \rho}
\end{equation}
in the Jordan frame. It is convenient to analyse the perturbation equations in the Einstein frame where $\rho= A(\phi) \tilde \rho$ implying that
\begin{equation}
\tilde \delta= \delta - \kappa_4\alpha_\phi \delta \phi
\end{equation}
where $\delta =\frac{\delta\rho}{\rho}$.
In the following we will calculate $\delta$ and $\delta \phi$.

In the Einstein frame, the absence of anisotropic stress implies that the metric can be written
\begin{equation}
ds^2_E= a^2 (-(1+2\Phi_N)d\eta^2 + (1-2\Phi_N)dx^2)
\end{equation}
in conformal time. Using $ds^2_J= A^2(\phi)ds_E^2$ and writing
\begin{equation}
ds^2_J= a^2 ((-(1+2\tilde \Psi)d\eta^2 + (1-2\tilde\Phi)dx^2)
\end{equation}
we find that
\begin{equation}
\tilde \Psi=  \Phi_N+ \frac{\alpha_\phi}{m_{\rm Pl}} \delta \phi,\ \ \tilde \Phi=  \Phi_N- \frac{\alpha_\phi}{m_{\rm Pl}} \delta \phi
\end{equation}
The Klein-Gordon equation becomes a relation between the density contrast and the field fluctuation
\begin{equation}
\delta\phi= -\alpha_{\phi} \frac{\kappa_4 a^2 \rho}{k^2 +a^2m_\phi^2}\delta
\end{equation}
in momentum space. We deduce that
\begin{equation}
\tilde \delta= (1+\alpha_\phi^2 \frac{3a^2 H^2}{k^2 +a^2m_\phi^2}) \delta
\end{equation}
In general, the correction term is small.

Using the Poisson equation
\begin{equation}
\Phi_N= -\frac{3a^2H^2}{2k^2}\delta
\end{equation}
we find that
\begin{equation}
\tilde \Psi= -\left [\frac{1+ \frac{k^2}{a^2m_\phi^2}(1+2\alpha_\phi^2)}{1+ \frac{k^2}{a^2m_\phi^2}}\right] \frac{3H^2a^2 \delta}{2k^2}
\end{equation}
and
\begin{equation}
\tilde \Phi= -\left [\frac{1+ \frac{k^2}{a^2m_\phi^2}(1-2\alpha_\phi^2)}{1+ \frac{k^2}{a^2m_\phi^2}}\right] \frac{3H^2a^2 \delta}{2k^2}
\end{equation}
This leads to the slip function\cite{Amendola:2007rr}
\begin{equation}
\tilde \gamma\equiv \frac{\tilde \Phi}{\tilde \Psi}= \frac{1+ \frac{k^2}{a^2m)\phi^2}(1-2\alpha_\phi^2)}{1+ \frac{k^2}{a^2m_\phi^2}(1+2\alpha_\phi^2)}
\end{equation}
Similarly one can identify a deviation from Poisson's law
\begin{equation}
k^2 \Psi= -\frac{3a^2H^2}{2}\tilde \mu \tilde \delta
\end{equation}
with
\begin{equation}
\tilde \mu= \left [\frac{1+ \frac{k^2}{a^2m_\phi^2}(1+2\alpha_\phi^2)}{1+ \frac{k^2}{a^2m_\phi^2}}\right] \frac{1}{(1+\alpha_\phi^2 \frac{3a^2 H^2}{k^2 +a^2m_\phi^2})}
\end{equation}
It is clear from these expressions that in chameleon theories,  the range of the chameleon force
\begin{equation}
\lambda_\phi=\frac{1}{m_\phi}
\end{equation}
 is crucial.
On large scales $k/a \ll m_\phi$, the deviations from general relativity disappear $\tilde \gamma=1,\ \tilde \mu=1$. This is not the case on scales smaller than $\lambda_c$, i.e.  $k/a\gg m_\phi$ where
\begin{equation}
\tilde \gamma =\frac{1-2\alpha_\phi^2}{1+2\alpha_\phi^2}, \ \ \tilde \mu= 1+2\alpha_\phi^2
\end{equation}
Notice that $\mu$ is tantamount  to the fact that Newton's constant for small objects is larger.
For $f(R)$ theories, we find that
\begin{equation}
\tilde \gamma= \frac{1}{2},\ \ \tilde \mu= \frac{4}{3}
\end{equation}
These are large deviations on small scales.

In fact, these modifications can be summarised in the equation governing $\delta$.  First define the divergence of the velocity fluid $\Theta= \partial_iv^i$ satisfying the Euler equation
\begin{equation}
\Theta' + {\cal H} \Theta= k^2 \tilde \psi
\end{equation}
where the conservation equation gives
\begin{equation}
\delta'= -\Theta
\end{equation}
Using these two equations we find:
\begin{equation}
\delta''+ {\cal H} \delta' -\frac{3}{2} {\cal H}^2 \frac{G_N(k)}{G_N}\delta=0
\end{equation}
in conformal time and in the linear regime. The scale dependent Newton constant is
\begin{equation}
G_{N}(k)=(1+ \frac{2\alpha_\phi^2}{1+\frac{a^2 m_\phi^2}{k^2}})G_N
\end{equation}
It interpolates between $G_{ N}$ on large scales $k/a \ll m_\phi$ and $G_{\rm N,eff}= (1+2\alpha_\phi^2) G_N$ on small scales $k/a\gg m_\phi$.
When the coupling $\alpha_\phi$ is constant, this equation can be integrated. When $k/a \ll m_\phi$, the growth is similar to the one in general relativity
\begin{equation}
\delta \sim a
\end{equation}
whereas gravity is  modified on small scales leading to an anomalous growth
\begin{equation}
\delta \sim a ^{\frac{\nu}{2}}
\end{equation}
where
\begin{equation}
\nu= \frac{-1+ \sqrt{1+ 24(1+2\alpha_\phi^2)}}{2}
\end{equation}
This implies that structures grow faster on small scales. For $f(R)$ models, we find that
\begin{equation}
\nu= \frac{-1+\sqrt {33}}{2}
\end{equation}
Using this result, one can evaluate the time dependence of $\tilde \Psi$ and $\tilde \Phi$. On top of being different, Newton's potentials $\tilde \Psi$ and $\tilde \Phi$ are now time dependent. This is important as this leads to the existence of an integrated Sachs-Wolfe effect in the Cosmic Microwave Background spectrum. The fact that peculiar velocities are sensitive to $\tilde \Psi$ while weak lensing is sensitive to $\tilde \Psi+ \tilde \Phi$ may also be used to disentangle the possible existence of a chameleon description of dark energy (of the $f(R)$ type for instance). Another important consequence is the fact that gravitational lensing is sensitive to $\tilde  \Psi +\tilde \Phi$, hence leading to an effect on weak lensing.

\section{Modified Gravity}

It is plausible that gravity may  be modified on large scales where it has never been tested experimentally. Such a modification could lead to the acceleration of the universe.

\subsection{ Massive Gravity}

The first and most natural modification of Einstein gravity is massive gravity. This is the gravitational analogue of Fermi's theory of massive gauge bosons. Like for spin 1 fields, the mass term breaks gauge invariance which is general covariance here
\begin{equation}\
g_{\mu\nu}\to g_{\mu\nu} + \partial_\mu \xi_\nu +\partial_\nu \xi_\mu
\end{equation}
It is simpler to study massive gravity in a Minkowski background by expanding
\begin{equation}
g_{\mu\nu}= \eta_{\mu\nu}+ h_{\mu\nu}
\end{equation}
where $h_{\mu\nu}$ is the graviton field.
The Lagrangian of massive gravity reads\cite{Rubakov:2008nh}
\begin{equation}
S=\frac{m_{\rm Pl}^2}{4} \int {\rm d^4} x \left [ ( \partial_\lambda h_{\mu\nu}\partial^\lambda h^{\mu\nu} -2\partial_\mu h^{\mu\nu} \partial_{\lambda} h^\lambda_\nu + 2 \partial_\mu h^{\mu\nu} \partial_{\nu} h^\lambda_\lambda- \partial_\mu h^\nu_\nu \partial^\mu h_\lambda^\lambda) + a h_{\mu\nu}h^{\mu\nu} + b(h_\mu^\mu)^2\right ]
\end{equation}
The first term is the expansion of the Einstein-Hilbert term up to second order while $a$ and $b$ parameterise the possible mass terms. Notice that we have  explicitly respected Lorentz invariance.

The analysis of ghosts in this model is easier using the Stuckelberg formalism.
Define
\begin{equation}
h_{\mu\nu}= \bar h_{\mu\nu} + \partial_\mu \phi_\nu + \partial_\nu \phi_\mu
\end{equation}
where under gauge variations
\begin{equation}
\bar h_{\mu\nu} \to \bar h_{\mu\nu} + \partial_\mu \xi_\nu + \partial_\nu \xi_\mu, \ \ \phi_\mu \to \phi_\mu - \xi_\mu
\end{equation}
The vector $\phi_\mu$ is the analogue of the longitudinal polarisation for massive vector fields.
This trick restores gauge invariance even in massive gravity. Now the Einstein-Hilbert part of the action is only a function of $\bar h_{\mu\nu}$ while the mass terms lead to kinetic terms for $\phi_\mu$. Considering only these kinetic terms, we find
\begin{equation}
{\cal L}_{\rm Stuckelberg}= \frac{m_{\rm Pl}^2}{2}\left [ a \partial_\mu\phi_\nu \partial^\mu \phi^\nu + (a+2b) (\partial_\mu \phi^\mu)^2\right ]
\end{equation}
This Lagrangian has no ghost provided it can be written up like in electrodynamics involving only $F_{\mu\nu}= \partial_\mu \phi_\nu -\partial_\nu\phi_\mu$. This is possible if
\begin{equation}
a=-b
\end{equation}
in which case
\begin{equation}
{\cal L}_{\rm Stuckelberg}=- \frac{m_{G}^2}{2}F_{\mu\nu}F^{\mu\nu}
\end{equation}
where we have introduce the mass
\begin{equation}
m^2_G= b m_{\rm Pl}^2
\end{equation}
which corresponds to the mass of the transverse traceless part of the graviton.
In summary, massive gravity has no ghost provided provided the mass term is of the special form
\begin{equation}
{\cal L}_{\rm massive}= \frac{m_{G}^2}{4}\left ( h_{\mu\nu}h^{\mu\nu} - (h^\mu_\mu)^2\right )
\end{equation}
called the Fierz-Pauli mass term.

A massless graviton has two polarisations while a massive graviton has five polarisations.
The gravitational field emitted by source of energy momentum tensor $T_{\mu\nu}$ is  in general relativity
\begin{equation}
h_{\mu\nu}=\frac{8\pi G_N}{p^2} \left [ T_{\mu\nu}-\frac{1}{2} \eta_{\mu\nu} T^\rho_\rho \right ]
\end{equation}
while in massive gravity
\begin{equation}
h_{\mu\nu}=\frac{8\pi G_N}{p^2+m_G^2} \left [ T_{\mu\nu}-\frac{1}{3} \eta_{\mu\nu} T^\rho_\rho \right ]
\end{equation}
Notice that the main difference is not the pole structure but the appearance of a $1/3$ instead of a $1/2$.
This implies that even in the vanishing limit $m_G\to 0$, massive gravity is not equivalent to massless gravity. This is due to the extra polarisations which are present in massive gravity. This major flaw is called the van Dam- Veltman- Zakharov discontinuity.
 At the non-linear level for a body of mass $M$, it has been argued by Vainshtein that there is radius
\begin{equation}
r_V= \left ( \frac{M}{m_{\rm Pl}^2 m_G^4}\right )^{1/5}
\end{equation}
below which massive gravity becomes non-linear and the vDVZ discontinuity is not applicable. Unfortunately, the ghost which was suppressed in a Minkowski background using the Fierz-Pauli choice  does not decouple anymore in a curved background, hence in  cosmology.

\subsection{DGP gravity}

Dvali-Gabadadze-Porrati gravity~\cite{Dvali:2000hr} is a modification of gravity at long distance using a 5d setting. It bears close resemblance  with massive gravity as we will see. The 5d action is
\begin{equation}
S= \int {\rm d^5 x} \sqrt{-g} \frac{R}{2\kappa_5^2} + \int {\rm d^4x} \sqrt{-g^{(4)}} \frac{R^{(4)}}{2 \kappa_4^2}
\end{equation}
where $g_{ab}$ is the 5d metric and $g^{(4)}_{\mu\nu}$ the induced metric on a 4d brane where matter lives.
This model is particularly popular as it leads to a naturally accelerating  brane cosmology. Consider the dynamics of the induced metric on the brane which we take to be of the flat FRW type. Using techniques which will not be described here, the Friedmann equation is modified to\cite{Deffayet:2001pu}
\begin{equation}
H^2\pm \frac{H}{r_0} =\frac{8\pi G_N}{3} \rho
\end{equation}
where matter lives on the brane and $8\pi G_N= \kappa_4^2$. The typical scale
\begin{equation}
r_0=\frac{\kappa_5^2}{2\kappa_4^2}
\end{equation}
is the new ingredient of the model.
The branch of solution with a $+$ in the Friedmann equation is such that,  for a low density $\rho \ll \kappa_4^2/\kappa_5^4$, we retrieve the usual Friedmann equation. On the contrary when the sign is $-$ called the self-accelerating branch, the Hubble rate tends to
\begin{equation}
H\to \frac{1}{r_0}
\end{equation}
asymptotically. Hence the cosmological constant leading to the acceleration of the universe is determined by the ratio of the 5d over the 4d gravitational couplings.

This success is marred by the properties of DGP gravity in the presence of a localised source on the brane
\begin{equation}
h_{\mu\nu}=\frac{8\pi G_N}{p^2+\frac{p}{r_0}} \left [ T_{\mu\nu}-\frac{1}{3} \eta_{\mu\nu} T^\mu_\nu \right ]
\end{equation}
At small scale, gravity is 4d and the interaction potential is in $1/r$ while at large scale $r\gg r_0$ gravity is 5d with a potential in $1/r^2$.
Hence gravity is modified at large scale by the opening of an extra dimension. Unfortunately, the $1/3$ term reminds us that gravity in the DGP model has five polarisations as  a massless graviton in 5d. We have seen that $r_0$ is of the order of the Hubble scale now in order to accommodate the acceleration of the universe. Hence gravity would have a wrong tensorial structure in the solar system and the DGP model would be ruled out. Fortunately, like in massive gravity, there is a Vainshtein scale
\begin{equation}
r_*= (r_S r_0^2)^{1/3},
\end{equation}
where $r_S= 2G_N M$ is the Schwarzchild radius of the source, below which DGP gravity behaves similarly  to general relativity.
On the other hand, DGP gravity suffers from the presence of ghosts on the self-accelerating branch preventing its use as a possible explanation of cosmic acceleration\cite{Charmousis:2006pn}. However, the DGP modification of gravity can be embedded in string theory\cite{Antoniadis:2002tr} with six extra dimensions. This provides an explicit regularisation of DGP gravity and cures it from the presence of ghosts.

\subsection{Ostrogradski's theorem}

We have just described two models of modified gravity suffering from the presence of ghosts. Gravity could be further modified by adding an arbitrary number of terms to the Einstein-Hilbert action involving higher order derivatives. In fact, there is a very strong result due to Ostrogradski which prevents one from doing so\cite{Woodard:2006nt}.

Consider a Lagrangian depending on a simple variable $q$ and let us focus on time dependent configurations only. The Lagrangian becomes
\begin{equation}
{\cal L}(q, q^{(1)}, \dots q^{(N)})
\end{equation}
where $q^{(j)}$ is the jth time derivative. The Euler-Lagrange equation is generalised to
\begin{equation}
\sum_{i=0}^N \left ( -\frac{d}{dt}\right )^i \frac{\partial L}{\partial q^{(i)}}=0
\end{equation}
Assuming that the Lagrangian is non-degenerate, one can express the Euler-Lagrange equation as
\begin{equation}
q^{(2N)}= G(q,\dots, q^{(2N-1)})
\end{equation}
Solving this equation requires $2N$ initial conditions. This implies that one must be able to define a $(2N)$ dimensional phase space.
The purpose of Ostrogradski's construction is to exhibit a conserved Hamiltonian in such a phase space, and to show that it is not bounded from below.

Choosing the canonical variables
\begin{equation}
Q_i\equiv q^{(i-1)}, \ \ P_i= \sum_{j=i}^N \left ( -\frac{d}{dt}\right )^{j-i} \frac{\partial L}{\partial q^{(j)}}
\end{equation}
one defines the Hamiltonian
\begin{equation}
H= \sum_{i=1}^N P_i q^{(i)} -L
\end{equation}
The Hamiltonian is conserved and generates the time evolution
\begin{equation}
\dot Q_i= \frac{\partial H}{\partial P_i},\ \ \dot P_i= - \frac{H}{\partial Q_i}
\end{equation}
If  the system is non-degenerate,  one can solve the equation
\begin{equation}
P_N= \frac{\partial L}{\partial q^{(N)}}
\end{equation}
and express
\begin{equation}
q^{(N)}= F( Q_1, \dots , Q_N, P_N).
\end{equation}
As a result, the phase space contains $2N$ coordinates $(Q_1,\dots , Q_N)$ and $(P_1, \dots , P_{N})$.
It is the clear that the Hamiltonian
\begin{equation}
H= P_1Q_2 + \dots+ P_{N-1} Q_N + P_N F- L(Q_1,\dots,Q_N, F)
\end{equation}
is unbounded from below in the $(N-1)$ direction $(P_1,\dots, P_{N-1})$. This is due to the presence of ghosts and is lethal for the theory.
A possible way out is to consider degenerate theories or gauge theories like ordinary gravity where constraints remove some of the ghost-like directions.

\section{Violation Of the Cosmological Principle}
We have just seen that constructing a modified theory of gravity is problematic. As it is also very hard to understand the properties of dark energy,
one may then doubt the validity of the cosmological  principle.
\subsection{Inhomogeneous universe}

The apparent acceleration of the universe could be due to a local inhomogeneity whose effect would mimic dark energy\cite{Biswas:2007gi}. This can be rendered rigorous by using the local Einstein equations in a patch around us and averaging over the volume of this patch. The averaged evolution of the universe does not only respond to the average density of matter as the Einstein equations are non-linear\cite{Buchert:2007ik}.

Let us assume that the universe is filled with a pressureless  fluid and that it can be foliated according to a global time coordinate
\begin{equation}
ds^2=-dt^2+ h_{ab}dx^adx^b
\end{equation}
where $x^a$ are spatial coordinates.
The expansion rate of the universe becomes a tensor
\begin{equation}
\Theta^a_b=\frac{1}{2} h^{ac}\dot h_{cb}
\end{equation}
Let us define $\Theta= \Theta^a_a$ and the  shear tensor $\sigma^a_b= \Theta^a_b-\frac{\Theta}{3} \delta^a_b$. The Einstein equations give the Friedmann equation
\begin{equation}
R^{(3)}+ \frac{2}{3}\Theta^2 -2 \sigma^2= 16\pi G_N \rho
\end{equation}
where $\sigma^2= \sigma_{ab}\sigma^{ab}$ and the  equation
\begin{equation}
R^{(3)}+\Theta^2 +\dot \Theta= 12\pi G_N \rho
\end{equation}
where $R^{(3)}$ is the curvature of $h_{ab}$.
These equations can be combined to give the Raychaudhury equation
\begin{equation}
\dot \Theta + \frac{1}{3} \Theta^2 + 2\sigma^2 +4\pi G_N \rho=0
\end{equation}
to describe the local geometry of the patch.

To describe global quantities, one needs to take the spatial average over a comoving volume $D$
\begin{equation}
<A>_D= \frac{1}{V_D} \int_D {\rm d^3} x \sqrt{h} A
\end{equation}
where $V_D=\int_D {\rm d^3} x \sqrt{h}$.
It is convenient to define the global scale factor
\begin{equation}
a_D^3=\frac{V_D}{V_{D,\rm {ini}}}
\end{equation}
compared to some initial volume. Using the commutation relation,
\begin{equation}
\frac{d<A>_D}{dt} - <\dot A>_D= <A\Theta>_D -<A>_D<\Theta>_D
\end{equation}
the averaged Friedmann and Raychaudhury equations become
\begin{equation}
H_D^2= \frac{8\pi G_N}{3} <\rho>_D -\frac{1}{6}(Q_D + <R^{(3)}>_D)
\end{equation}
and
\begin{equation}
\frac{\ddot a_D}{a_D}=-  \frac{4\pi G_N}{3} <\rho>_D +\frac{1}{3} Q_D
\end{equation}
where $H_D= \dot a_D/a_D$ and
\begin{equation}
Q_D=  \frac{2}{3}( <\Theta^2>_D- <\Theta>_D^2) -2 \sigma^2.
\end{equation}
As expected, the evolution equations for the averaged cosmology receive corrections terms due to the
non-linearities of the Einstein equations.
Surprisingly, these equations are equivalent to the ones obtained with a dark energy fluid $\phi_D$ whose potential reads\cite{Buchert:2006ya}
\begin{equation}
V(\phi_D)=-\frac{<R^{(3)}>_D}{24\pi G_N}
\end{equation}
and
\begin{equation}
\dot \phi_D^2=V(\phi_D)-\frac{Q_D}{8\pi G_N}
\end{equation}
Of course this result is very intriguing as it states that dark energy could be the result of the local dynamics of our local patch of universe.
To go further, one needs to calculate the averages in a particular model.

\subsection{Tolman-Bondi universe}

The CMB gives a very precise indication that the universe is isotropic at large. If acceleration is due to a local inhomogeneity in our surrounding patch, the phenomenological constraints coming from SN Ia and the CMB are quite tight\cite{Alexander:2007xx}. First of all, outside our patch, the Hubble rate should be as low as $h\sim 0.45$. Moreover one must accomodate a patch reaching as far as the first supernovae $z\sim 0.05$ implying the existence of a large void of size between $160/h $Mpc and $250/h$ Mpc. Inside this underdensity, the density contrast must be as large as $\delta\sim -0.4$. All these constraints can be met using special solutions of Einstein equations based on the metric
\begin{equation}
ds^2=-dt^2 + \frac{R'^2(r,t)}{1-k(r)r^2}dr^2 + R^2(r,t)(d\theta^2 + \sin^2\theta d\phi^2)
\end{equation}
known as a Tolman-Bondi universe. Notice that these space-times are isotropic but inhomogeneous. They generalise the FRW solutions and represent an explicit violation of the Copernican principle: the observers have a special place in the universe. The Friedmann  equation  can be expressed in terms of
\begin{equation}
\frac{2G_NM(r)}{R(r,t)}-k(r)r^2= \dot R(r,t)^2
\end{equation}
where the comoving mass function $M(r)$ inside a comoving patch of radius $r$ is time independent and  is related to the density inside the patch
\begin{equation}
M(r)= 4\pi \int_0^r {\rm d}r'\rho(r',t) R'(r',t) R^2(r',t)
\end{equation}
The function $k(r)$ is the local curvature inside the patch.
Let us consider a patch where the curvature is negative. The solution of the Friedmann equation is
\begin{equation}
R(r,t)= -\frac{G_N M(r)}{k(r) r^2} (\cosh \eta -1)
\end{equation}
where time is given by
\begin{equation}
t-t_0(r)= \frac{G_N M(r)}{{(-k(r)r^2)}^{3/2}} (\sinh \eta -\eta)
\end{equation}
The Tolman-Bondi universes have two arbitrary functions: the curvature $k(r)$ and the initial time $t_0(r)$ interpreted as the time of the big bang
at the location $r$.
The volume inside the comoving sphere is given by
\begin{equation}
V_D= 4\pi\int_0^{r_D} \frac{R'(r,t)R^2(r,t)}{\sqrt{1-k(r)r^2}} dr
\end{equation}
It has been shown that the averaged acceleration parameter\cite{Paranjape:2009zu}
\begin{equation}
q_D= -\frac{\ddot a_D a_D}{\dot a_D^2}
\end{equation}
is negative when
\begin{equation}
I_{kr^2}>\frac{I_{kr}^2}{3I_k}
\end{equation}
where
\begin{equation}
I_k= 2\pi \int_0^{r_D} {\rm d} r \frac{\sqrt{-kr^2}(-kr^2)'}{\sqrt{1-kr^2}},\
I_{kr}= 4\pi \int_0^{r_D} {\rm d} r \frac{(-kr^3)'}{\sqrt{1-kr^2}}
\end{equation}
and
\begin{equation}
I_{kr^2}= 4\pi \int_0^{r_D} {\rm d} r \frac{(r^3 \sqrt{-k})'}{\sqrt{1-kr^2}}
\end{equation}
In particular choosing
\begin{equation}
k(r)=-\frac{1}{1+r^a},\ 0<a<2
\end{equation}
leads to an effective acceleration.
It would be very interesting to have more general results and a correspondence with the equivalent dark energy description.

\section{Conclusions}

These lecture notes have presented some of the issues related to the observation of cosmic acceleration. The list of topics which have been covered is rather small compared to the breadth of the existing literature. Four points of view have been  commonly advocated. The first one which is favoured by some astronomers is that a pure cosmological constant is enough to explain all the physics behind cosmic acceleration. I hope I have managed to convey the idea  that a sheer cosmological constant is not entirely satisfactory and calls for a deeper explanation. One of the possibilities could be that the value of the cosmological constant has an anthropic origin. Another explanation could be dark energy although it requires a huge cancelation too and explains only  the observation of a tiny residual cosmological constant. In the class of dark energy models, axion-like models are the best motivated ones from the particle physics point of view. Another alternative could be modified gravity. As I have recalled, this is a tortuous route strewn with difficulties such as the existence of ghosts. In view of its relevance, the challenge of building an appropriate infrared modification of general relativity is worth pursuing. Finally the violation of the Copernican principle and the role of inhomogeneities are very hot topics which may hold the secret of cosmic acceleration.

\bibliography{references}

\providecommand{\href}[2]{#2}\begingroup\raggedright\begin{thebibliography}{10}

\bibitem{Perlmutter:1998np}
{\bf Supernova Cosmology Project} Collaboration, S.~Perlmutter {\em et~al.},
  {\it Measurements of omega and lambda from 42 high-redshift supernovae},
  {\em Astrophys. J.} {\bf 517} (1999) 565--586,
  [\href{http://xxx.lanl.gov/abs/astro-ph/9812133}{{\tt astro-ph/9812133}}].

\bibitem{Riess:1998cb}
{\bf Supernova Search Team} Collaboration, A.~G. Riess {\em et~al.}, {\it
  Observational evidence from supernovae for an accelerating universe and a
  cosmological constant},  {\em Astron. J.} {\bf 116} (1998) 1009--1038,
  [\href{http://xxx.lanl.gov/abs/astro-ph/9805201}{{\tt astro-ph/9805201}}].

\bibitem{Lyth:1998xn}
D.~H. Lyth and A.~Riotto, {\it {Particle physics models of inflation and the
  cosmological density perturbation}},  {\em Phys. Rept.} {\bf 314} (1999)
  1--146, [\href{http://xxx.lanl.gov/abs/hep-ph/9807278}{{\tt
  hep-ph/9807278}}].

\bibitem{Trodden:2004st}
M.~Trodden and S.~M. Carroll, {\it {TASI lectures: Introduction to cosmology}},
   \href{http://xxx.lanl.gov/abs/astro-ph/0401547}{{\tt astro-ph/0401547}}.

\bibitem{Will:2005va}
C.~M. Will, {\it {The confrontation between general relativity and
  experiment}},  {\em Living Rev. Rel.} {\bf 9} (2005) 3,
  [\href{http://xxx.lanl.gov/abs/gr-qc/0510072}{{\tt gr-qc/0510072}}].

\bibitem{2008arXiv0803.0547K}
E.~{Komatsu}, J.~{Dunkley}, M.~R. {Nolta}, C.~L. {Bennett}, B.~{Gold},
  G.~{Hinshaw}, N.~{Jarosik}, D.~{Larson}, M.~{Limon}, L.~{Page}, D.~N.
  {Spergel}, M.~{Halpern}, R.~S. {Hill}, A.~{Kogut}, S.~S. {Meyer}, G.~S.
  {Tucker}, J.~L. {Weiland}, E.~{Wollack}, and E.~L. {Wright}, {\it {Five-Year
  Wilkinson Microwave Anisotropy Probe (WMAP) Observations: Cosmological
  Interpretation}},  {\em ArXiv e-prints} {\bf 803} (Mar., 2008)
  [\href{http://xxx.lanl.gov/abs/0803.0547}{{\tt 0803.0547}}].

\bibitem{Weinberg:1988cp}
S.~Weinberg, {\it The cosmological constant problem},  {\em Rev. Mod. Phys.}
  {\bf 61} (1989) 1--23.

\bibitem{Dvali:2000hr}
G.~R. Dvali, G.~Gabadadze, and M.~Porrati, {\it {4D gravity on a brane in 5D
  Minkowski space}},  {\em Phys. Lett.} {\bf B485} (2000) 208--214,
  [\href{http://xxx.lanl.gov/abs/hep-th/0005016}{{\tt hep-th/0005016}}].

\bibitem{Shaposhnikov:2008xb}
M.~Shaposhnikov and D.~Zenhausern, {\it {Scale invariance, unimodular gravity
  and dark energy}},  {\em Phys. Lett.} {\bf B671} (2009) 187--192,
  [\href{http://xxx.lanl.gov/abs/0809.3395}{{\tt 0809.3395}}].

\bibitem{Caldwell:2009ix}
R.~R. Caldwell and M.~Kamionkowski, {\it {The Physics of Cosmic Acceleration}},
   \href{http://xxx.lanl.gov/abs/0903.0866}{{\tt 0903.0866}}.

\bibitem{Silvestri:2009hh}
A.~Silvestri and M.~Trodden, {\it {Approaches to Understanding Cosmic
  Acceleration}},  {\em Rept. Prog. Phys.} {\bf 72} (2009) 096901,
  [\href{http://xxx.lanl.gov/abs/0904.0024}{{\tt 0904.0024}}].

\bibitem{Copeland:2006wr}
E.~J. Copeland, M.~Sami, and S.~Tsujikawa, {\it {Dynamics of dark energy}},
  {\em Int. J. Mod. Phys.} {\bf D15} (2006) 1753--1936,
  [\href{http://xxx.lanl.gov/abs/hep-th/0603057}{{\tt hep-th/0603057}}].

\bibitem{Burgess:2004ib}
C.~P. Burgess, {\it {Towards a natural theory of dark energy: Supersymmetric
  large extra dimensions}},  {\em AIP Conf. Proc.} {\bf 743} (2005) 417--449,
  [\href{http://xxx.lanl.gov/abs/hep-th/0411140}{{\tt hep-th/0411140}}].

\bibitem{Burgess:2005wu}
C.~P. Burgess, {\it {Supersymmetric large extra dimensions and the cosmological
  constant problem}},  \href{http://xxx.lanl.gov/abs/hep-th/0510123}{{\tt
  hep-th/0510123}}.

\bibitem{Bousso:2000xa}
R.~Bousso and J.~Polchinski, {\it {Quantization of four-form fluxes and
  dynamical neutralization of the cosmological constant}},  {\em JHEP} {\bf 06}
  (2000) 006, [\href{http://xxx.lanl.gov/abs/hep-th/0004134}{{\tt
  hep-th/0004134}}].

\bibitem{Martin:1997ns}
S.~P. Martin, {\it {A Supersymmetry Primer}},
  \href{http://xxx.lanl.gov/abs/hep-ph/9709356}{{\tt hep-ph/9709356}}.

\bibitem{Kamenshchik:2001cp}
A.~Y. Kamenshchik, U.~Moschella, and V.~Pasquier, {\it An alternative to
  quintessence},  {\em Phys. Lett.} {\bf B511} (2001) 265--268,
  [\href{http://xxx.lanl.gov/abs/gr-qc/0103004}{{\tt gr-qc/0103004}}].

\bibitem{Bean:2003ae}
R.~Bean and O.~Dore, {\it {Are Chaplygin gases serious contenders to the dark
  energy throne?}},  {\em Phys. Rev.} {\bf D68} (2003) 023515,
  [\href{http://xxx.lanl.gov/abs/astro-ph/0301308}{{\tt astro-ph/0301308}}].

\bibitem{Caldwell:2003vq}
R.~R. Caldwell, M.~Kamionkowski, and N.~N. Weinberg, {\it {Phantom Energy and
  Cosmic Doomsday}},  {\em Phys. Rev. Lett.} {\bf 91} (2003) 071301,
  [\href{http://xxx.lanl.gov/abs/astro-ph/0302506}{{\tt astro-ph/0302506}}].

\bibitem{Ratra:1987rm}
B.~Ratra and P.~J.~E. Peebles, {\it {Cosmological Consequences of a Rolling
  Homogeneous Scalar Field}},  {\em Phys. Rev.} {\bf D37} (1988) 3406.

\bibitem{Ferreira:1997hj}
P.~G. Ferreira and M.~Joyce, {\it {Cosmology with a Primordial Scaling Field}},
   {\em Phys. Rev.} {\bf D58} (1998) 023503,
  [\href{http://xxx.lanl.gov/abs/astro-ph/9711102}{{\tt astro-ph/9711102}}].

\bibitem{Wetterich:1987fm}
C.~Wetterich, {\it {Cosmology and the Fate of Dilatation Symmetry}},  {\em
  Nucl. Phys.} {\bf B302} (1988) 668.

\bibitem{Albrecht:1999rm}
A.~J. Albrecht and C.~Skordis, {\it {Phenomenology of a realistic accelerating
  universe using only Planck-scale physics}},  {\em Phys. Rev. Lett.} {\bf 84}
  (2000) 2076--2079, [\href{http://xxx.lanl.gov/abs/astro-ph/9908085}{{\tt
  astro-ph/9908085}}].

\bibitem{Steinhardt:1999nw}
P.~J. Steinhardt, L.-M. Wang, and I.~Zlatev, {\it {Cosmological tracking
  solutions}},  {\em Phys. Rev.} {\bf D59} (1999) 123504,
  [\href{http://xxx.lanl.gov/abs/astro-ph/9812313}{{\tt astro-ph/9812313}}].

\bibitem{Amendola:1999er}
L.~Amendola, {\it {Coupled quintessence}},  {\em Phys. Rev.} {\bf D62} (2000)
  043511, [\href{http://xxx.lanl.gov/abs/astro-ph/9908023}{{\tt
  astro-ph/9908023}}].

\bibitem{Cline:2003gs}
J.~M. Cline, S.~Jeon, and G.~D. Moore, {\it {The phantom menaced: Constraints
  on low-energy effective ghosts}},  {\em Phys. Rev.} {\bf D70} (2004) 043543,
  [\href{http://xxx.lanl.gov/abs/hep-ph/0311312}{{\tt hep-ph/0311312}}].

\bibitem{Kolda:1998wq}
C.~F. Kolda and D.~H. Lyth, {\it {Quintessential difficulties}},  {\em Phys.
  Lett.} {\bf B458} (1999) 197--201,
  [\href{http://xxx.lanl.gov/abs/hep-ph/9811375}{{\tt hep-ph/9811375}}].

\bibitem{Barbieri:2005gj}
R.~Barbieri, L.~J. Hall, S.~J. Oliver, and A.~Strumia, {\it {Dark energy and
  right-handed neutrinos}},  {\em Phys. Lett.} {\bf B625} (2005) 189--195,
  [\href{http://xxx.lanl.gov/abs/hep-ph/0505124}{{\tt hep-ph/0505124}}].

\bibitem{Damour:1992we}
T.~Damour and G.~Esposito-Farese, {\it {Tensor multiscalar theories of
  gravitation}},  {\em Class. Quant. Grav.} {\bf 9} (1992) 2093--2176.

\bibitem{Bertotti:2003rm}
B.~Bertotti, L.~Iess, and P.~Tortora, {\it {A test of general relativity using
  radio links with the Cassini spacecraft}},  {\em Nature} {\bf 425} (2003)
  374.

\bibitem{Brax:2004ym}
P.~Brax, C.~van~de Bruck, and A.~C. Davis, {\it {Is the radion a chameleon?}},
  {\em JCAP} {\bf 0411} (2004) 004,
  [\href{http://xxx.lanl.gov/abs/astro-ph/0408464}{{\tt astro-ph/0408464}}].

\bibitem{Brax:2009kd}
P.~Brax, C.~van~de Bruck, J.~Martin, and A.-C. Davis, {\it {Decoupling Dark
  Energy from Matter}},  {\em JCAP} {\bf 0909} (2009) 032,
  [\href{http://xxx.lanl.gov/abs/0904.3471}{{\tt 0904.3471}}].

\bibitem{Choi:1999xn}
K.~Choi, {\it {String or M theory axion as a quintessence}},  {\em Phys. Rev.}
  {\bf D62} (2000) 043509, [\href{http://xxx.lanl.gov/abs/hep-ph/9902292}{{\tt
  hep-ph/9902292}}].

\bibitem{Khoury:2003rn}
J.~Khoury and A.~Weltman, {\it {Chameleon cosmology}},  {\em Phys. Rev.} {\bf
  D69} (2004) 044026, [\href{http://xxx.lanl.gov/abs/astro-ph/0309411}{{\tt
  astro-ph/0309411}}].

\bibitem{Brax:2004qh}
P.~Brax, C.~van~de Bruck, A.-C. Davis, J.~Khoury, and A.~Weltman, {\it
  {Detecting dark energy in orbit: The cosmological chameleon}},  {\em Phys.
  Rev.} {\bf D70} (2004) 123518,
  [\href{http://xxx.lanl.gov/abs/astro-ph/0408415}{{\tt astro-ph/0408415}}].

\bibitem{Adelberger:2004ct}
E.~G. Adelberger, {\it {Short-range tests of the gravitational inverse square
  law}}, . Prepared for 3rd Meeting on CPT and Lorentz Symmetry (CPT 04),
  Bloomington, Indiana, 4-7 Aug 2004.

\bibitem{Sotiriou:2008rp}
T.~P. Sotiriou and V.~Faraoni, {\it {f(R) Theories Of Gravity}},
  \href{http://xxx.lanl.gov/abs/0805.1726}{{\tt 0805.1726}}.

\bibitem{Amendola:2007nt}
L.~Amendola and S.~Tsujikawa, {\it {Phantom crossing, equation-of-state
  singularities, and local gravity constraints in $f(R)$ models}},  {\em Phys.
  Lett.} {\bf B660} (2008) 125--132,
  [\href{http://xxx.lanl.gov/abs/0705.0396}{{\tt 0705.0396}}].

\bibitem{Brax:2008hh}
P.~Brax, C.~van~de Bruck, A.-C. Davis, and D.~J. Shaw, {\it {f(R) Gravity and
  Chameleon Theories}},  {\em Phys. Rev.} {\bf D78} (2008) 104021,
  [\href{http://xxx.lanl.gov/abs/0806.3415}{{\tt 0806.3415}}].

\bibitem{Amendola:2007rr}
L.~Amendola, M.~Kunz, and D.~Sapone, {\it {Measuring the dark side (with weak
  lensing)}},  {\em JCAP} {\bf 0804} (2008) 013,
  [\href{http://xxx.lanl.gov/abs/0704.2421}{{\tt 0704.2421}}].

\bibitem{Rubakov:2008nh}
V.~A. Rubakov and P.~G. Tinyakov, {\it {Infrared-modified gravities and massive
  gravitons}},  {\em Phys. Usp.} {\bf 51} (2008) 759--792,
  [\href{http://xxx.lanl.gov/abs/0802.4379}{{\tt 0802.4379}}].

\bibitem{Deffayet:2001pu}
C.~Deffayet, G.~R. Dvali, and G.~Gabadadze, {\it {Accelerated universe from
  gravity leaking to extra dimensions}},  {\em Phys. Rev.} {\bf D65} (2002)
  044023, [\href{http://xxx.lanl.gov/abs/astro-ph/0105068}{{\tt
  astro-ph/0105068}}].

\bibitem{Charmousis:2006pn}
C.~Charmousis, R.~Gregory, N.~Kaloper, and A.~Padilla, {\it {DGP
  specteroscopy}},  {\em JHEP} {\bf 10} (2006) 066,
  [\href{http://xxx.lanl.gov/abs/hep-th/0604086}{{\tt hep-th/0604086}}].

\bibitem{Antoniadis:2002tr}
I.~Antoniadis, R.~Minasian, and P.~Vanhove, {\it {Non-compact Calabi-Yau
  manifolds and localized gravity}},  {\em Nucl. Phys.} {\bf B648} (2003)
  69--93, [\href{http://xxx.lanl.gov/abs/hep-th/0209030}{{\tt
  hep-th/0209030}}].

\bibitem{Woodard:2006nt}
R.~P. Woodard, {\it {Avoiding dark energy with 1/R modifications of gravity}},
  {\em Lect. Notes Phys.} {\bf 720} (2007) 403--433,
  [\href{http://xxx.lanl.gov/abs/astro-ph/0601672}{{\tt astro-ph/0601672}}].

\bibitem{Biswas:2007gi}
T.~Biswas and A.~Notari, {\it {Swiss-Cheese Inhomogeneous Cosmology and the
  Dark Energy Problem}},  {\em JCAP} {\bf 0806} (2008) 021,
  [\href{http://xxx.lanl.gov/abs/astro-ph/0702555}{{\tt astro-ph/0702555}}].

\bibitem{Buchert:2007ik}
T.~Buchert, {\it {Dark Energy from Structure - A Status Report}},  {\em Gen.
  Rel. Grav.} {\bf 40} (2008) 467--527,
  [\href{http://xxx.lanl.gov/abs/0707.2153}{{\tt 0707.2153}}].

\bibitem{Buchert:2006ya}
T.~Buchert, J.~Larena, and J.-M. Alimi, {\it {Correspondence between
  kinematical backreaction and scalar field cosmologies: The 'morphon field'}},
   {\em Class. Quant. Grav.} {\bf 23} (2006) 6379--6408,
  [\href{http://xxx.lanl.gov/abs/gr-qc/0606020}{{\tt gr-qc/0606020}}].

\bibitem{Alexander:2007xx}
S.~Alexander, T.~Biswas, A.~Notari, and D.~Vaid, {\it {Local Void vs Dark
  Energy: Confrontation with WMAP and Type Ia Supernovae}},  {\em JCAP} {\bf
  0909} (2009) 025, [\href{http://xxx.lanl.gov/abs/0712.0370}{{\tt
  0712.0370}}].

\bibitem{Paranjape:2009zu}
A.~Paranjape, {\it {The Averaging Problem in Cosmology}},
  \href{http://xxx.lanl.gov/abs/0906.3165}{{\tt 0906.3165}}.

\end{thebibliography}\endgroup

\end{document}